\documentstyle[amsfonts,aps]{revtex}

\begin{document}
\title{Switching Boundary Conditions in the Many-Body Diffusion Algorithm}
\author{F. Luczak, F.~Brosens 
\thanks{Senior Research Associate of the FWO-Vlaanderen.}%
%
and J.T. Devreese 
\thanks{Also at the Universiteit Antwerpen (RUCA) and 
       Technische Universiteit Eindhoven, The Netherlands.}%
%
}
\address{Departement Natuurkunde, Universiteit Antwerpen (UIA),\\
Universiteitsplein 1, B-2610 Antwerpen}
\author{L. F. Lemmens}
\address{Departement Natuurkunde, Universiteit Antwerpen (RUCA),\\
Groenenborgerlaan 171, B-2020 Antwerpen}
\date{January 27, 1998}
\maketitle

\begin{abstract}
In this paper we show how the transposition, the basic operation of the
permutation group, can be taken into account in a diffusion process of
identical particles. Whereas in an earlier approach the method was applied
to systems in which the potential is invariant under interchanging the
Cartesian components of the particle coordinates, this condition on the
potential is avoided here. In general, the potential introduces a switching
of the boundary conditions of the walkers. [These transitions modelled by a
continuous--time Markov chain generate sample paths for the propagator as a
Feynman--Kac functional . A few examples , including harmonic fermions with
an anharmonic interaction, and the ground-state energy of ortho-helium are
studied to elucidate the theoretical discussion and to illustrate the
feasibility of a sign-problem-free implementation scheme for the recently
developed many-body diffusion approach.]
\end{abstract}

\section{Introduction}

In a previous paper \cite{LBDL97}, the present authors have shown that the
many-body diffusion algorithm (MBDA) allows for a sign-problem-free
simulation of excited antisymmetric states of interacting harmonic
oscillators. The basic idea behind this approach has been to split the
Euclidean-time propagator into a sum of independent propagators which remain
positive on the appropriate state space \cite{BDL94,BDL95,BDL96,BDLer97}.
Each of the elementary propagators could be individually sampled as a
diffusion process for distinguishable particles on a state space with
absorbing or reflecting boundaries.

In the present paper, the many-body diffusion formalism and its
implementation are generalized to allow for the construction of
interdependent diffusion processes. In this approach the propagator that
governs the Euclidean time evolution is split into two parts: the kinetic
part is used to describe the evolution of the system with well-defined
boundary conditions, the potential is used for branching and killing and
rules transitions from one set of boundary conditions to another set. The
evolution is given by a Brownian motion or a random walk, but the boundary
conditions can change during the simulation according to a continuous-time
Markov process. The transition rates of this process partly or completely
derive from the potential which determines also the killing or branching
process.

This principle is illustrated with some test models. A limited number of
degrees of freedom is considered to avoid large computation times and to
allow for a visual control on the probability densities that are generated.
The presented algorithm does not integrate smoothly with the previous one
for harmonic fermions, in the sense that for this algorithm the interchange
of two identical particles is fully under control but in combination with
the even permutations it is far from trivial if more than two particles of
parallel spin are present.

In the next section, the theoretical background of the method is discussed.
In Sec. III, details on the diffusion in a reduced state space are provided.
In Sec. IV, a sign-problem-free implementation is developed and applied to
two toy models to illustrate the concepts. Subsequently, anisotropic
harmonic fermions are studied avoiding normal coordinates in order to
elucidate the role of the potential for the transitions between different
sets of boundary conditions. Finally, the MBDA is employed to calculate the
ground-state energy of ortho-helium. Some concluding remarks are presented
in Sec. V.

\section{A sign-problem-free approach}

In this section, we describe the construction of a Feynman-Kac functional
equipped with a process to generate the sample paths over which the
exponential containing the interaction has to be averaged. This construction
is based on the many body diffusion approach introduced in \cite{BDL96} and
illustrated in \cite{LBDL97}. The procedure is sign-problem-free in the
strict sense, i.e. expectation values in the Feynman-Kac functional are
obtained using transition probabilities in a Wiener-Poisson space. A
mathematical account on the foundations of these compound processes can be
found in \cite{WilRo87}. The Wiener space is the support for the diffusion
whereas a multivariate Poisson process counts the switches of the boundary
condition as will be explained below.

\subsection{Diffusion on a Domain}

Let $\overline{r}$ be a $3N$-dimensional point in the configuration space $%
{\Bbb R}^{3N}$ of $N$ distinguishable particles and assume that the
propagator for this system is given by: 
\begin{equation}
K_{D}[\overline{r}_{f},\tau ;\overline{r}_{i},0]=\sum_{k}\left( \psi _{k}(%
\overline{r}_{i}),\exp (-\tau H)\,\psi _{k}(\overline{r}_{f})\right) \quad ,
\label{PropD}
\end{equation}
where $H$ describes the evolution in Euclidean time of the N particles. The
subscript $D$ denotes that at this stage the particles are still considered
to be distinguishable. This propagator can be easily written as a
Feynman-Kac functional over all paths starting in $\overline{r}_{i}\,$ and
ending in $\overline{r}_{f}$ a Euclidean time lapse $\tau $ later. The paths
are constructed by a $3N$-dimensional Brownian motion $\left\{ \overline{R}%
\left( \tau \right) ;\tau \geq 0\right\} $ with variance $\sigma ^{2}=\hbar
\tau /m$%
\begin{equation}
K_{D}[\overline{r}_{f},\tau ;\overline{r}_{i},0]=E_{\overline{r}_{i}}\left[
I_{\left( \overline{R}\left( \tau \right) =\overline{r}_{f}\right) }\exp
\!\left( -\frac{1}{\hbar }\int\limits_{0}^{\tau }V\left( \overline{R}%
(\varsigma )\right) d\varsigma \right) \right] \quad .  \label{FKD}
\end{equation}
All integration paths in (\ref{FKD}) start in $\overline{r}_{i}$ as
indicated by the averaging index $E_{\overline{r}_{i}}$ and satisfy the
condition $\overline{R}\left( \tau \right) =\overline{r}_{f}$ as denoted by
the indicator $I_{\left( \overline{R}\left( \tau \right) =\overline{r}%
_{f}\right) }$. Writing the propagator as an average over a Brownian motion
is equivalent to giving the evolution equation and the initial condition: 
\begin{equation}
\frac{\partial }{\partial \tau }K_{D}[\overline{r}_{f}{\bf ,}\tau ;\overline{
r}_{i},0]=\left( \frac{\hbar }{2m}\nabla _{\overline{r}{\bf \,}}^{2}-\frac{1%
}{\hbar }V(\overline{r}_{f})\right) K_{D}[\overline{r}_{f}{\bf ,}\tau ;%
\overline{r}_{i},0]\quad ,  \label{Esch}
\end{equation}
\begin{equation}
\lim_{\tau \downarrow 0}K_{D}[\overline{r}_{f},\tau ;\overline{r}%
_{i},0]=\delta \left( \overline{r}_{f}-\overline{r}_{i}\,\right) \quad .
\label{InC}
\end{equation}
If a propagator containing {\sl two indistinguishable }particles $j$ and $k$
has to be derived from (\ref{PropD}), a projection on the symmetric
(anti-symmetric) irreducible representation of the permutation group has to
be made for bosons (fermions): 
\begin{equation}
K_{I}[\overline{r}_{f},\tau ;\overline{r}_{i},0]=\frac{1}{2!}\sum_{P}\xi
^{P}K_{D}[P\overline{r}_{f},\tau ;\overline{r}_{i},0]\quad ,  \label{PropI}
\end{equation}
where $P$ represents the permutations of the particles $j$,$k$ and $\xi =\pm
1$ is chosen according to the particle statistics under consideration. The
evolution equation for the projected propagator is given by the same eq. (%
\ref{Esch}), because $\nabla _{\overline{r}{\bf \,}}^{2}$ as well as the
potential $V\left( \overline{r}\right) $ are invariant under all the
permutations $P$. The initial condition obtained from the projection (\ref
{PropI}) by taking the limit $\tau \downarrow 0$ is not the initial
condition that is usually studied if one wants to solve a partial
differential equation using probabilistic methods.\cite{ChuZha95} In order
to obtain an initial condition which is given by 
\begin{equation}
\lim_{\tau \downarrow 0}K_{I}[\overline{r}_{f},\tau ;\overline{r}%
_{i},0]=\delta \left( \overline{r}_{f}-\overline{r}_{i}\right) \;\text{with}
\quad \overline{r}_{f}{\bf ,}\overline{r}_{i}\in D_{2}^{3}\otimes {\Bbb R}
^{3\left( N-2\right) }\quad ,  \label{InCid}
\end{equation}
one has to restrict the configuration space ${\Bbb R}^{3N}$ to a domain $%
D_{2}^{3}\otimes {\Bbb R}^{3\left( N-2\right) }$. In this domain the
identical-particle coordinates $\vec{r}_{j},\vec{r}_{k}$ are linearly
ordered. In $D_{2},$ for example, the $x$--coordinates are ordered as
follows: $x_{j}\geq x_{k}.$ Introducing a domain implies that one has to
specify the boundary conditions; the specification appropriate for bosons or
fermions moving freely or interacting harmonically has been proposed in \cite
{BDL96} and analyzed in \cite{LBDL97}.

\subsubsection{Boundary conditions}

The ordering particles in a three--dimensional space can be performed by
ordering the particles with respect to each of the three independent
Cartesian directions. This option has been chosen in the previous
investigations and led \cite{LBDL97,BDL94,BDL95,BDL96} to the construction
of an ordered state-space $D$. The {\sl symmetric} irreducible
representation implies four combinations $l=0,1,2,3.$ By definition, the ($%
l=0)$--boundary condition causes reflection at the boundary $\partial D$ in
every direction $x,y,z$, whereas for $l=1,2,3$ the boundary $\partial D$
reflects in $x,y,z$-direction and absorbs in the two other two Cartesian
directions. For fermions, $l=0$ means absorption at $\partial D$ in all
three directions, whereas for $l=1$ absorption occurs in the $x$-direction
and reflection in the $y$- and $z$-direction. Similarly, $l=2$ and $l=3$
correspond to the appropriate absorption and reflection conditions with
respect to the $y$- and $z$-direction. It is clear that in order to study
the evolution of the system in Euclidean time one has to give not only the
starting point in the domain $D_2^3$ but also the boundary conditions $l$.
The propagator has to contain the information that a state, initially
characterized by $\overline{r}_i$ and $l, $ will end after a Euclidean time
lapse $\tau $ in $\overline{r}_f$ with boundary conditions $\hat{l}$. The
evolution in Euclidean time of this propagator can be studied from its
extension to ${\Bbb R}^{3N}$, which gives 
\begin{equation}
K_I[\overline{r}_f,l{\bf ,}\tau ;\overline{r}_i,\hat{l},0]=\left( \frac 1{2!}%
\right) ^3\sum_{P_x}\xi _l^{P_x}\sum_{P_y}\xi _l^{P_y}\sum_{P_z}\xi
_l^{P_z}K_D[P_xP_yP_z\,\overline{r}_f,\tau ;\overline{r}_i,\hat{l},0]\quad ,
\label{PropIBc}
\end{equation}
\noindent%
%
where $P_x,P_y,P_z$ symbolize the permutations of the $x,y,z$-coordinates of
the two indistinguishable particles $j$ and $k$ and where we imposed the
initial condition 
\begin{equation}
\lim_{\tau \downarrow 0}K_I[\overline{r}_f,l{\bf ,}\tau ;\overline{r}_i,\hat{
l},0]=\delta _{l{\bf ,}\hat{l}}\;\delta (\overline{r}_f-\overline{r}_i)\quad
,\quad \overline{r}_f{\bf ,}\overline{r}_i\in D_2^3\otimes {\Bbb R}^{3\left(
N-2\right) }\quad .  \label{InCidbc}
\end{equation}
The boundary condition $l$ determines the parity $\xi _l^{P_x},\xi
_l^{P_y},\xi _l^{P_z}$ of the permutations $P_x,P_y,P_z$.

\subsubsection{Evolution equations}

A potential $V\left( \overline{r}\right) $ whose decomposition contains only
one-dimensional irreducible$^{{}}$ representations with respect to
permutations $Q_{x},Q_{y},Q_{z}$ of the coordinates is straightforward to
analyze. Expanding the potential energy in the possible symmetry
combinations, described above, one may write: 
\begin{equation}
V(\overline{r})=\sum_{l^{\prime }=0}^{3}\left( \frac{1}{2!}\right)
^{3}\sum_{Q_{x}}\xi _{l^{\prime }}^{Q_{x}}\sum_{Q_{y}}\xi _{l^{\prime
}}^{Q_{y}}\sum_{Q_{z}}\xi _{l^{\prime }}^{Q_{z}}V(Q_{x}Q_{y}Q_{z}\,\overline{
r}\,)\quad .  \label{Irlb}
\end{equation}
Consider now the following decomposition 
\begin{eqnarray*}
\chi &=&\left( \frac{1}{2!}\right)
^{3}\sum_{P_{x}}\sum_{P_{y}}\sum_{P_{z}}\xi _{l}^{P_{x}}\xi _{l}^{P_{y}}\xi
_{l}^{P_{z}}\,V(P_{x}P_{y}P_{z}\,\overline{r}_{f})\;K_{D}[P_{x}P_{y}P_{z}\,%
\overline{r}_{f}\,,\tau ;\overline{r}_{i},\hat{l},0] \\
&=&\left( \frac{1}{2!}\right) ^{6}\sum_{l^{\prime
}}\sum_{P_{x}Q_{x}}\sum_{P_{y}Q_{y}}\sum_{P_{z}Q_{z}}\xi _{l}^{P_{x}}\xi
_{l^{\prime }}^{Q_{x}}\xi _{l}^{P_{y}}\xi _{l^{\prime }}^{Q_{y}}\xi
_{l}^{P_{z}}\xi _{l^{\prime }}^{Q_{z}}\;V(Q_{x}P_{x}Q_{y}P_{y}Q_{z}P_{z}\,%
\overline{r}_{f})\;K_{D}[P_{x}P_{y}P_{z}\,\overline{r}_{f}\,,\tau ;\overline{
r}_{i},\hat{l},0]\,.
\end{eqnarray*}
Introducing the permutations $S_{i}=Q_{i}P_{i}$ with $i=x,y,z$, the quantity 
$\chi $ can be written as: 
\[
\chi =\left( \frac{1}{2!}\right)
^{6}\sum_{P_{x}S_{x}}\sum_{P_{y}S_{y}}\sum_{P_{z}S_{z}}\sum_{l^{\prime }}\xi
_{l}^{P_{x}}\xi _{l^{\prime }}^{S_{x}P_{x}^{-1}}\xi _{l}^{P_{y}}\xi
_{l^{\prime }}^{S_{y}P_{y}^{-1}}\xi _{l}^{Pz}\xi _{l^{\prime
}}^{S_{z}P_{z}^{-1}}\,V(S_{x}S_{y}S_{z}\,\overline{r}_{f})%
\;K_{D}[P_{x}P_{y}P_{z}\,\overline{r}_{f}\,,\tau ;\overline{r}_{i},\hat{l}
\,,0]\,. 
\]
The permutation $P_{i}$ and its inverse $P_{i}^{-1}$have the same parity and
therefore $\xi _{l^{\prime }}^{S_{i}P_{i}^{-1}}=\xi _{l^{\prime
}}^{S_{i}}\xi _{l^{\prime }}^{P_{i}}$. We furthermore define $\xi _{\bar{l}%
}^{P_{i}}\equiv \xi _{l}^{P_{i}}\xi _{l^{\prime }}^{P_{i}}$. This leads
immediately to: 
\begin{equation}
\chi =\sum_{\overline{l}}V_{l\overline{l}}(\overline{r}_{f}){\bf \,}K_{I}[%
\overline{r}_{f},\overline{l}{\bf ,}\tau ;\overline{r}_{i},\hat{l}\,,0]\;,
\label{def}
\end{equation}
\noindent%
%
where we introduced the notation: 
\begin{equation}
V_{l\overline{l}}(\overline{r}_{f}){\bf \,}=\left( \frac{1}{2!}\right)
^{3}\sum_{S_{x}}\sum_{S_{y}}\sum_{S_{z}}\xi _{l}^{S_{x}}\xi _{l}^{S_{y}}\xi
_{l}^{S_{z}}\,V(S_{x}S_{y}S_{z}\,\overline{r}_{f})\,\xi _{\bar{l}%
}^{S_{x}}\xi _{\bar{l}}^{S_{y}}\xi _{\bar{l}}^{S_{z}}\quad .  \label{Rate}
\end{equation}
Using eq. (\ref{Esch}) for the propagator (\ref{PropIBc}) and the result
obtained in (\ref{def}), the evolution equation becomes: 
\begin{equation}
\frac{\partial }{\partial \tau }\,K_{I}[\overline{r}_{f},l{\bf ,}\tau ;%
\overline{r}_{i},\hat{l}\,,0]=\frac{\hbar }{2m}\,\nabla _{\overline{r}{\bf \,%
}}^{2}\,K_{I}[\overline{r}_{f},l{\bf ,}\tau ;\overline{r}_{i},\hat{l},0]-%
\frac{1}{\hbar }\sum_{\bar{l}}V_{l\overline{l}}(\overline{r}_{f}){\bf \;}
K_{I}[\overline{r}_{f},\bar{l}{\bf ,}\tau ;\overline{r}_{i},\hat{l},0]\,,
\label{Ediju}
\end{equation}
where use has been made of the invariance of $\nabla _{\overline{r}{\bf \,}
}^{2}$ with respect to permutations of the coordinates $x,y,z$ of the
particle positions.

The study of the non-diagonal interaction contribution (\ref{Ediju}) is the
main purpose of the present section. In some cases, these non-diagonal terms
are zero. For instance, if $V\left( \overline{r}\right) $ can be written as $%
V_s\left( \overline{r}\right) =V\left( \overline{x}\right) +V(\overline{y}
)+V\left( \overline{z}\right) ,$ the invariance under the permutations of
the particles implies invariance under permutations of the coordinates,
which in turn then implies that $V_{ll^{\prime }}(\overline{r})=\delta
_{ll^{\prime }}V(\overline{r}).$ For harmonic interactions along the
principal axes the many-body potential can be written as such a sum. The
evolution equation is then diagonal in the quantum numbers $l$, meaning that
during the evolution the boundary conditions do not change. The
computational feasibility and efficiency of this type of evolution equation
has been demonstrated in \cite{LBDL97}.

\subsubsection{Decomposition of the Potential}

The potential $V(\vec{r}_j;\vec{r}_k)\equiv V(\vec{r}_1,\ldots ,\vec{r}
_j,\ldots ,\vec{r}_k,\ldots ,\vec{r}_N)$ is invariant under the permutation
of any two particles, and in particular for interchanging $\vec{r}_j$ with $%
\vec{r}_k$ 
\begin{equation}
V(\,\vec{r}_j;\,\vec{r}_k)=V(\,\vec{r}_k;\,\vec{r}_j)\;.
\end{equation}
For motion in $2D$, one can decompose the potential as: 
\begin{equation}
V(x_j,y_j;x_k,y_k)=V_{\text{bb}}(x_j,y_j;x_k,y_k)+V_{\text{ff}%
}(x_j,y_j;x_k,y_k)\;,
\end{equation}
with 
\[
V_{\text{bb}}(x_j,y_j;x_k,y_k)=\frac 14\,\left[
V(x_j,y_j;x_k,y_k)+V(x_k,y_j;x_j,y_k)+V(x_k,y_k;x_j,y_j)+V(x_j,y_k;x_k,y_j)%
\right] 
\]
and 
\[
V_{\text{ff}}(x_j,y_j;x_k,y_k)=\frac 14\,\left[
V(x_j,y_j;x_k,y_k)-V(x_k,y_j;x_j,y_k)+V(x_k,y_k;x_j,y_j)-V(x_j,y_k;x_k,y_j)%
\right] \;. 
\]
Introducing the operator $\sigma _x,\sigma _y$ for interchanging the $x,y$
-coordinates, and the operator $e_x,e_y$ for leaving the $x,y$-coordinates
unchanged, the decomposition of the potential can be written as: 
\[
V_{\text{bb}}(x_j,y_j;x_k,y_k)=\frac 14\left( e_x+\sigma _x\right) \left(
e_y+\sigma _y\right) V(x_j,y_j;x_k,y_k)\;. 
\]
For motion in $3D$ the same principle can be used leading to the
decomposition: 
\begin{eqnarray}
V(x_j,y_j,z_j;x_k,y_k,z_k) &=&V_{\text{bbb}}(x_j,y_j,z_j;x_k,y_k,z_k)+V_{%
\text{bff}}(x_j,y_j,z_j;x_k,y_k,z_k) \\
&&+V_{\text{fbf}}(x_j,y_j,z_j;x_k,y_k,z_k)+V_{\text{ffb}
}(x_j,y_j,z_j;x_k,y_k,z_k)\;,  \nonumber
\end{eqnarray}
where typically $V_{\text{fbf}}(x_j,y_j,z_j;x_{k,}y_k,z_k)$ is given by: 
\begin{equation}
V_{\text{fbf}}(x_j,y_j,z_j;x_k,y_k,z_k)=\frac 18\left( e_x-\sigma _x\right)
\left( e_y+\sigma _y\right) \left( e_z-\sigma _z\right)
V(x_j,y_j,z_j;x_k,y_k,z_k)
\end{equation}
and analogous definitions hold for the other matrix elements. It is
interesting to note the similarity between this projection and quaternions: 
\begin{equation}
1_{xyz}=\frac 18\left[ 
\begin{array}{c}
\left( e_x+\sigma _x\right) \left( e_y+\sigma _y\right) \left( e_z+\sigma
_z\right) +\left( e_x+\sigma _x\right) \left( e_y-\sigma _y\right) \left(
e_z-\sigma _z\right) \\ 
+\left( e_x-\sigma _x\right) \left( e_y+\sigma _y\right) \left( e_z-\sigma
_z\right) +\left( e_x-\sigma _x\right) \left( e_y-\sigma _y\right) \left(
e_z+\sigma _z\right)
\end{array}
\right] \;.
\end{equation}
Using this projection the formal decomposition (\ref{Rate}) is achieved
explicitly.

\subsection{The Feynman-Kac functional}

Given that the propagator indeed satisfies the evolution equation (\ref
{Ediju}), it remains to be shown that a process can be found to generate the
sample paths and a Feynman-Kac functional.

\subsubsection{Positivity and the process}

To start the construction of the generator of the process consider the
following identity: 
\begin{eqnarray}
\sum_{\bar{l}}V_{l\overline{l}}\left( \,\overline{r}\right) &=&\left( \frac{
1}{2!}\right) ^{3}\sum_{\bar{l}}\sum_{S_{x}}\sum_{S_{y}}\sum_{S_{z}}\xi
_{l}^{S_{x}}\xi _{l}^{S_{y}}\xi _{l}^{S_{z}}V(S_{x}S_{y}S_{z}\,\,\overline{r}
)\,\xi _{\bar{l}}^{S_{x}}\xi _{\bar{l}}^{S_{y}}\xi _{\bar{l}}^{S_{z}}
\label{ident} \\
&=&V\left( \overline{r}\right) \quad .  \nonumber
\end{eqnarray}
This is a direct consequence of eq. (\ref{Irlb}). Assume further, that the
domain $D_{2}^{3}$ is chosen in such a way that $V_{l\overline{l}}\left( 
\overline{r}\right) \leq 0$ for $l\neq \bar{l}$ . This allows to rewrite eq.
(\ref{Ediju}) as follows: 
\begin{eqnarray}
\frac{\partial }{\partial \tau }K_{I}[\overline{r}_{f},l{\bf ,}\tau ;%
\overline{r}_{i},\hat{l},0] &=&\frac{\hbar }{2m}\,\nabla _{\,\overline{r}%
}^{2}\,K_{I}[\overline{r}_{f},l{\bf ,}\tau ;\overline{r}_{i},\hat{l}\,,0]-%
\frac{1}{\hbar }\,V(\overline{r}_{f})\,K_{I}[\overline{r}_{f},l{\bf ,}\tau ;%
\overline{r}_{i},\hat{l},0]  \label{evcmdf} \\
&&-W_{l}\left( \overline{r}_{f}\right) K_{I}[\overline{r}_{f},l{\bf ,}\tau ;%
\overline{r}_{i},\hat{l}\,,0]+\sum_{\bar{l}}W_{l\overline{l}}\left( 
\overline{r}_{f}\right) K_{I}[\overline{r}_{f},\bar{l}{\bf ,}\tau ;\overline{
r}_{i},\hat{l}\,,0]\,,  \nonumber
\end{eqnarray}
where $W_{l\overline{l}}\left( \overline{r}_{f}\right) $ denotes the
non-diagonal part $-\frac{1}{\hbar }V_{l\overline{l}}\left( \overline{r}%
_{f}\right) $ if $l\neq \bar{l}$: 
\begin{equation}
W_{l\overline{l}}\left( \overline{r}_{f}\right) =-\frac{1}{\hbar }V_{l%
\overline{l}}\left( \overline{r}_{f}\right) \,(1-\delta _{l,\overline{l}
})\quad ,  \label{ndrate}
\end{equation}
and the quantity $W_{l}\left( \overline{r}_{f}\right) $ stands for: 
\begin{equation}
W_{l}\left( \overline{r}_{f}\right) =\sum_{\overline{l}}W_{l\overline{l}
}\left( \overline{r}_{f}\right)  \label{drate}
\end{equation}
In this situation, the $W_{l\overline{l}}\left( \overline{r}_{f}\right) $
can be interpreted as the rate that a walker in the position $\overline{r}
_{f}{\bf \,}\,$will change its boundary condition from $l$ to $\overline{l}$
. This means that the following terms of eq. (\ref{evcmdf}) generate a
diffusion together with a Markov process on the four possible boundary
conditions: 
\begin{eqnarray}
\frac{\partial }{\partial \tau }K_{I}^{0}[\overline{r}_{f},l{\bf ,}\tau ;%
\overline{r}_{i},\hat{l},0] &=&\frac{\hbar }{2m}\,\nabla _{\,\overline{r}%
}^{2}\,K_{I}^{0}[\overline{r}_{f},l{\bf ,}\tau ;\overline{r}_{i},\hat{l}%
\,,0]+\sum_{\bar{l}}W_{l\overline{l}}\left( \overline{r}_{f}\right)
K_{I}^{0}[\overline{r}_{f},\bar{l}{\bf ,}\tau ;\overline{r}_{i},\hat{l}
\,,0]\,  \label{evcmpr} \\
&&-W_{l}\left( \overline{r}_{f}\right) K_{I}^{0}[\overline{r}_{f},l{\bf ,}
\tau ;\overline{r}_{i},\hat{l}\,,0],  \nonumber
\end{eqnarray}
where $K_{I}^{0}[\overline{r}_{f},l{\bf ,}\tau ;\overline{r}_{i},\hat{l}%
\,,0] $ satisfies a backward Kolmogorov equation, and there for it is the
transition probability of a compound stochastic process \cite{WilRo87} that
generates the sample paths in our state space.

\subsubsection{Positivity and the functional}

Above, we assumed that the domain $D_{2}^{3}$ can be chosen such that $V_{l%
\overline{l}}\left( \overline{r}\right) \leq 0$ for $l\neq \overline{l}$.
This assumption is necessary, because otherwise the introduced $W$--matrix
would not represent a continuous-time Markov process. From its construction
it is clear that it depends on the potential. Therefore a general analysis
for any potential is beyond the scope of the present communication. Assuming
that the system under consideration has a potential with this property (or
can be transformed by a unitary transformation to a system with such a
potential), the evolution (\ref{evcmpr}) describes a diffusion triggering a
continuous--time Markov chain that swithces boundary conditions in a
Wiener--Poisson space. The switches do not induce discontinuities in the
sample paths but allow for a change in boundary condition according to the
rates defined by the $W$--matrix. The sign condition on the projections of
the potential is necessary for the interpretation of the non diagonal
elements as rates. The evolution equation based on this assumption describes
a stochastic process in the strict mathematical sense. It can therefore be
simulated sign-problem-free. The propagator $K_{I}[\overline{r}_{f},${\bf $l$%
}${\bf ,}\tau ;\overline{r}_{i},\hat{l},0]$ with the initial condition (\ref
{InCidbc}) can be written as a Feynman-Kac functional $\;$%
\begin{equation}
K_{I}[\overline{r}_{f},l{\bf ,}\tau ;\overline{r}_{i},\hat{l},0]=E_{%
\overline{r}_{i},\hat{l}}\,\left[ I_{\left( \overline{R}\left( \tau \right) =%
\overline{r}_{f},L_{\tau }=l\right) }\exp \!\left( -\frac{1}{\hbar }\sum_{%
\overline{l}}\int\limits_{0}^{\tau }V_{L_{\varsigma }\overline{l}}\left( 
\overline{R}(\varsigma )\right) d\varsigma \right) \right] \,,  \label{piind}
\end{equation}
where use has been made of the identity (\ref{ident}) to make explicit that
the potential is diagonal in the label $l.$

The sample paths are generated by a $3N$-dimensional Brownian motion $%
\left\{ \overline{R}\left( \tau \right) ;\tau \geq 0\right\} $ with $\sigma
^{2}=\hbar \tau /m$ and absorbing and reflecting boundary conditions as
defined by the index $l$. After the transition, the boundary conditions are
adapted to agree with those from the Markov chain $\left\{ L_{\tau };\tau
\geq 0\right\} $ where the (normalized) quantities $V_{l\overline{l}}$ then
determine the boundary-condition transition rates $\overline{l}\rightarrow l$
. Repeated application of this procedure for all the subprocesses starting
with boundary conditions $\overline{l}$ generates the sample paths of the
diffusion process of $N-2$ distinguishable and two indistinguishable
particles in three dimensions. The stochastic integration over the
Feynman-Kac functional is then done by killing and branching.

\section{Diffusion on A Reduced State Space $\widehat{D}$}

The positivity of the rates $W_{l\overline{l}}(\overline{r}_{f})$ is a
necessary ingredient to guarantee the interpretation of the Euclidean time
Schr\"{o}dinger equation as a compound diffusion equation with killing and
branching in a Wiener-Poisson space. For some potentials, the condition of
negative definite $V_{l\overline{l}}(\overline{r}_{f})$ may not be satisfied
on the state space $D_{2}^{3}\otimes {\Bbb R}^{3\left( N-2\right) }.$ In
these cases, the definition of an alternative (reduced) state space is
required in order to find purely positive transition rates $W_{l\overline{l}
}(\overline{r}_{f})$. The symmetry of the potential effects the structure of
the adapted state space, which therewith is specific for certain classes of
applications. To illustrate the construction principle, the reduced state
space $\widehat{D}_{2}^{3}\otimes {\Bbb R}^{3\left( N-2\right) }$ will be
considered, where 
\begin{equation}
\widehat{D}_{2}^{3}\equiv \left\{ \left. \left(
x_{j},x_{k},y_{j},y_{k},z_{j},z_{k}\right) ^{T}\right| x_{j}\geq
|x_{k}|,y_{j}\geq |y_{k}|,z_{j}\geq |z_{k}|\right\} \;.  \label{reduced}
\end{equation}
\noindent This choice will be of relevance for the probabilistic sampling of
the real ground-state wave function of ortho-helium. The procedures for the
construction of an ordered state space $x_{j}\geq x_{k}$, $y_{j}\geq y_{k}$, 
$z_{j}\geq z_{k}$, reported in \cite{BDL94,BDL95,BDL96}, alone do not induce
the ordering on $\widehat{D}_{2}^{3},$ and an additional set of boundary
conditions must be supplied to impose the desired ordering. In that
framework, the separation of the multi-dimensional diffusion process into
one-dimensional diffusion processes will again prove useful.

\subsection{Free Diffusion of Indistinguishable Particles on $\widehat{D}
_2^1 $}

The free diffusion process of two indistinguishable particles on a line has
been introduced in \cite{BDL94,BDL95,BDL96} as the diffusion process of two
distinguishable particles on the ordered state space $D_{2}^{1}$ with
absorption (for fermions) or reflection (for bosons) at the boundary $%
\partial D_{2}^{1}$. The construction of the free diffusion process on the
reduced state space $\widehat{D}_{2}^{1}$ requires the extension of the
stochastic process. (Anti)symmetrization again plays the major role in that
approach: the diffusion process of free indistinguishable particles on the
state space $D_{2}^{1}$ can be mapped onto $\widehat{D}_{2}^{1}$ by the
construction of reflecting and absorbing boundaries. The starting point for
the derivation of the required processes is the separation of the state
space into two non-intersecting subspaces $\widehat{D}_{2}^{1}$ and $%
\widehat{P}_{x}\widehat{D}_{2}^{1}$ of equal volume: $D_{2}^{1}=$ $\widehat{D%
}_{2}^{1}\cup \widehat{P}_{x}\widehat{D}_{2}^{1}$. The mapping between both
subspaces is realized by the permutation operator $\widehat{P}_{x}$ which
models reflection at $x_{1}=-x_{2}$, i.e.: $\widehat{P}%
_{x}(x_{1},x_{2})^{T}=(-x_{2},-x_{1})^{T}$. Then, the transition probability
density $\rho _{\text{f,b}}(\overline{x},\tau ;\overline{x}^{\,\prime })$
that the system of two free fermions/bosons evolves from $\overline{x}%
^{\,\prime }$ to $\overline{x}$ during the Euclidean time interval $\tau $,
can be written as 
\begin{equation}
\rho _{\text{f,b}}(\overline{x},\tau ;\overline{x}^{\,\prime })=\frac{1}{2}%
\left( \rho _{\text{f,b}}^{\text{f}}(\overline{x},\tau ;\overline{x}%
^{\,\prime })+\rho _{\text{f,b}}^{\text{b}}(\overline{x},\tau ;\overline{x}%
^{\,\prime })\right) \quad ,  \label{def2}
\end{equation}
where we defined 
\[
\rho _{\text{f,b}}^{\text{f}}(\overline{x},\tau ;\overline{x}^{\,\prime
})=\rho _{\text{f,b}}(\overline{x},\tau ;\overline{x}^{\,\prime })-\rho _{%
\text{f,b}}(\widehat{P}_{x}\overline{x},\tau ;\overline{x}^{\,\prime })\quad
,\quad \rho _{\text{f,b}}^{\text{b}}(\overline{x},\tau ;\overline{x}%
^{\,\prime })=\rho _{\text{f,b}}(\overline{x},\tau ;\overline{x}^{\,\prime
})+\rho _{\text{f,b}}(\widehat{P}_{x}\overline{x},\tau ;\overline{x}%
^{\,\prime })\;. 
\]
Each of the transition probability densities $\rho _{\text{f,b}}(\overline{x}
,\tau ;\overline{x}^{\,\prime })$ consists of two density matrices, one---$%
\rho _{\text{f,b}}^{\text{b}}(\overline{x},\tau ;\overline{x}^{\,\prime })$%
---for which the boundary $x_{1}=-x_{2}$ acts reflecting, and one---$\rho _{%
\text{f,b}}^{\text{f}}(\overline{x},\tau ;\overline{x}^{\,\prime })$---which
induces absorption at $x_{1}=-x_{2}$. With an appropriate initial choice $%
\overline{x}^{\,\prime }\in \widehat{D}_{2}^{1}$, free fermion/boson
diffusion to any point $\overline{x}\in D_{2}^{1}$---and thus also to any $%
\overline{x}\in {\Bbb R}^{2}$---can be projected to the reduced state space $%
\widehat{D}_{2}^{1}$ by means of reflecting and absorbing state-space
boundaries. Each of the elementary density matrices $\rho _{\text{f,b}}^{%
\text{b}}(\overline{x},\tau ;\overline{x}^{\,\prime })$ and $\rho _{\text{f,b%
}}^{\text{f}}(\overline{x},\tau ;\overline{x}^{\,\prime })$ constitutes a
diffusion process with the initial condition 
\begin{equation}
\lim_{\tau \downarrow 0}\,\rho _{\text{f,b}}^{\text{b}}(\overline{x},\tau ;%
\overline{x}^{\,\prime })=\lim_{\tau \downarrow 0}\,\rho _{\text{f,b}}^{%
\text{f}}(\overline{x},\tau ;\overline{x}^{\,\prime })=\delta (\overline{x}-%
\overline{x}^{\,\prime })\quad .
\end{equation}
To proof this statement, $\rho _{\text{f}}^{\text{b}}(\overline{x},\tau ;%
\overline{x}^{\,\prime })$ and $\rho _{\text{f}}^{\text{f}}(\overline{x}%
,\tau ;\overline{x}^{\,\prime })$ must be shown to satisfy probability
conservation and the semigroup property. Both properties are derived taking
into account the Markov properties for the free fermion/boson diffusion
density $\rho _{\text{f,b}}(\overline{x},\tau ;\overline{x}^{\,\prime })$
defined on the state space $D_{2}^{1}$ with absorbing/reflecting boundaries.
Probability conservation for the two-boson density $\rho _{\text{b}}^{\text{b%
}}(\overline{x},\tau ;\overline{x}^{\,\prime })$ follows from 
\[
\int_{\widehat{D}_{2}^{1}}d\overline{x}\,\rho _{\text{b}}^{\text{b}}(%
\overline{x},\tau ;\overline{x}^{\,\prime })=\frac{1}{2}\int_{\widehat{D}%
_{2}^{1}}d\overline{x}\,\sum_{\widehat{P}_{x}}\rho _{\text{b}}^{\text{b}}(%
\widehat{P}_{x}\overline{x},\tau ;\overline{x}^{\,\prime })=\frac{1}{2}%
\int_{D_{2}^{1}}d\overline{x}\;\rho _{\text{b}}^{\text{b}}(\overline{x},\tau
;\overline{x}^{\,\prime })=\frac{1}{2}\int_{D_{2}^{1}}d\overline{x}\;\left(
\rho _{\text{b}}(\overline{x},\tau ;\overline{x}^{\,\prime })+\rho _{\text{b}
}(\widehat{P}_{x}\overline{x},\tau ;\overline{x}^{\,\prime })\right) =1\,. 
\]
As argued in \cite{BDL94}, the free fermion diffusion density $\rho _{\text{f%
}}(\overline{x},\tau ;\overline{x}^{\,\prime })$ is not automatically
normalized on the state space $D_{2}^{1}$. In contrast, normalization of the
free fermion diffusion density $\rho _{\text{f}}(\overline{x},\tau ;%
\overline{x}^{\,\prime })$ requires the introduction of the absorbing
boundary state \cite{BDL94} which comprises all transitions to the outside
of the state space $D_{2}^{1}$. Also in the present analysis, the
construction of a boundary state \cite{SKOHO} \cite{ChuZha95} ensures the
normalization of the probability densities $\rho _{\text{b}}^{\text{f}}(%
\overline{x},\tau ;\overline{x}^{\,\prime })$, $\rho _{\text{f}}^{\text{b}}(%
\overline{x},\tau ;\overline{x}^{\,\prime })$ and $\rho _{\text{f}}^{\text{f}
}(\overline{x},\tau ;\overline{x}^{\,\prime })$. The density $\rho _{\text{b}
}^{\text{f}}(\overline{x},\tau ;\overline{x}^{\,\prime })$, for instance,
associates an absorbing boundary $x_{1}=-x_{2}$ to bear in mind the
probability for distinguishable-particle propagation to $x_{1}\leq -x_{2}.$
The semigroup properties of $\rho _{\text{f}}^{\text{f}}(\overline{x},\tau ;%
\overline{x}^{\,^{\prime }})$, $\rho _{\text{f}}^{\text{b}}(\overline{x}
,\tau ;\overline{x}^{\,^{\prime }})$, $\rho _{\text{b}}^{\text{f}}(\overline{
x},\tau ;\overline{x}^{\,^{\prime }})$ and $\rho _{\text{b}}^{\text{b}}(%
\overline{x},\tau ;\overline{x}^{\,^{\prime }})$ are derived by unfolding to
the state space $D_{2}^{1}$. Using their symmetry properties, they can be
retraced to the semigroup properties of $\rho _{\text{f}}(\overline{x},\tau ;%
\overline{x}^{\,^{\prime }})$ and $\rho _{\text{b}}(\overline{x},\tau ;%
\overline{x}^{\,^{\prime }})$. E.g. 
\begin{eqnarray*}
\int_{\widehat{D}_{2}^{1}}d\overline{x}\,\rho _{\text{f}}^{\text{f}}(%
\overline{x}^{\,\prime },\tau ;\overline{x})\,\rho _{\text{f}}^{\text{f}}(%
\overline{x},\sigma ;\overline{x}^{\,\prime \prime }) &=&\frac{1}{2}\int_{%
\widehat{D}_{2}^{1}}d\overline{x}\sum_{\widehat{P}_{x}}\,\rho _{\text{f}}^{%
\text{f}}(\overline{x}^{\,\prime },\tau ;\widehat{P}_{x}\overline{x})\,\rho
_{\text{f}}^{\text{f}}(\widehat{P}_{x}\overline{x},\sigma ;\overline{x}%
^{\,\prime \prime }) \\
&=&\frac{1}{2}\int_{D_{2}^{1}}d\overline{x}\,\left( \rho _{\text{f}}(%
\overline{x}^{\,\prime },\tau ;\overline{x})-\rho _{\text{f}}(\overline{x}%
^{\,\prime },\tau ;\widehat{P}_{x}\overline{x})\right) \left( \,\rho _{\text{
f}}(\overline{x},\sigma ;\overline{x}^{\,\prime \prime })-\rho _{\text{f}}(%
\widehat{P}_{x}\overline{x},\sigma ;\overline{x}^{\,\prime \prime })\right)
\\
&=&\int_{D_{2}^{1}}d\overline{x}\,\left( \rho _{\text{f}}(\overline{x}%
^{\,\prime },\tau ;\overline{x})\,\rho _{\text{f}}(\overline{x},\sigma ;%
\overline{x}^{\,\prime \prime })-\rho _{\text{f}}(\overline{x}^{\,\prime
},\tau ;\overline{x})\,\rho _{\text{f}}(\overline{x},\sigma ;\widehat{P}_{x}%
\overline{x}^{\,\prime \prime })\right) \\
&=&\rho _{\text{f}}(\overline{x}^{\,\prime },\tau +\sigma ;\overline{x}
^{\,\prime \prime })\,-\rho _{\text{f}}(\overline{x}^{\,\prime },\tau
+\sigma ;\widehat{P}_{x}\overline{x}^{\,\prime \prime })\,=\rho _{\text{f}}^{%
\text{f}}(\overline{x}^{\,\prime },\tau +\sigma ;\overline{x}^{\,\prime
\prime })\,\;.
\end{eqnarray*}
As a consequence, the free diffusion process $\widehat{X}_{\text{f,b}}(\tau
) $ of two indistinguishable particles on a line is characterized by the
combination of two adapted Wiener processes defined on the reduced state
space $\widehat{D}_{2}^{1}$: 
\begin{eqnarray*}
\widehat{X}_{\text{b}}(\tau ) &=&\frac{1}{2}\left( \widehat{X}_{\text{b}}^{%
\text{b}}(\tau )\,+\widehat{X}_{\text{b}}^{\text{f}}(\tau )\right) \;\text{%
for bosons, \quad and\qquad } \\
\,\widehat{X}_{\text{f}}(\tau ) &=&\frac{1}{2}\left( \widehat{X}_{\text{f}}^{%
\text{b}}(\tau )\,+\widehat{X}_{\text{f}}^{\text{f}}(\tau )\right) \;\,\,%
\text{for fermions }.
\end{eqnarray*}
Since $\rho _{\text{b,f}}^{\text{b}}(\overline{x},\tau ;\overline{x}%
^{\,\prime })$ and $\rho _{\text{b,f}}^{\text{f}}(\overline{x},\tau ;%
\overline{x}^{\,\prime })$ are orthogonal, e.g. 
\begin{eqnarray*}
\int_{D_{2}^{1}}d\overline{x}\,\rho _{\text{f}}^{\text{f}}(\overline{x}%
^{\,\prime },\tau ;\overline{x})\,\rho _{\text{f}}^{\text{b}}(\overline{x}%
,\tau ;\overline{x}^{\,\prime \prime }) &=&\int_{\widehat{D}_{2}^{1}}d%
\overline{x}\,\rho _{\text{f}}^{\text{f}}(\overline{x}^{\,\prime },\tau ;%
\overline{x})\,\rho _{\text{f}}^{\text{b}}(\overline{x},\tau ;\overline{x}%
^{\,\prime \prime })+\int_{\widehat{P}_{x}\widehat{D}_{2}^{1}}d\overline{x}%
\,\rho _{\text{f}}^{\text{f}}(\overline{x}^{\,\prime },\tau ;\overline{x}%
)\,\rho _{\text{f}}^{\text{b}}(\overline{x},\tau ;\overline{x}^{\,\prime
\prime }) \\
&=&\int_{\widehat{D}_{2}^{1}}d\overline{x}\,\left[ \rho _{\text{f}}^{\text{f}%
}(\overline{x}^{\,\prime },\tau ;\overline{x})\,\rho _{\text{f}}^{\text{b}}(%
\overline{x},\tau ;\overline{x}^{\,\prime \prime })-\rho _{\text{f}}^{\text{f%
}}(\overline{x}^{\,\prime },\tau ;\overline{x})\,\rho _{\text{f}}^{\text{b}}(%
\overline{x},\tau ;\overline{x}^{\,\prime \prime })\right] =0\,,
\end{eqnarray*}
each of the densities $\rho _{\text{b,f}}^{\text{b}}(\overline{x},\tau ;%
\overline{x}^{\,^{\prime }})$, $\rho _{\text{b,f}}^{\text{f}}(\overline{x}%
,\tau ;\overline{x}^{\,^{\prime }})$ evolves independently and generates the
corresponding Brownian motion. We have used in the upper and the lower
indices the same letter to characterize the boundary condition, this is
customarily done in supersymmetric models \cite{CoKH95}, it should be noted
that these letters for b in the upper indices indicate wether there is
absorption or reflection on the boundary introduced by the (broken) symmetry
requirements for the potential and they are independent of the f or b in the
under indices introduced by the statistics of the particles.

In the sampling procedures, each of the processes $\widehat{X}_{\text{f,b}}^{%
\text{f,b}}(\tau )$ is constructed from the diffusion process $X_{\text{d}%
}(\tau )$ of two distinguishable particles taking into account the
corresponding boundary conditions: the process $\widehat{X}_{\text{b}}^{%
\text{f}}(\tau )$, for instance, means reflection at the hypersurface $%
x_{1}=x_{2}$ and absorption at $x_{1}=-x_{2}$. For each of the composite
processes, absorption at the boundary defines a characteristic {\sl first
exit time} $\tau _{\!\partial \widehat{D}_{2}^{1}}$ \cite{BDL96} which
specifies the adapted stochastic process, e.g. 
\[
\widehat{X}_{\text{f}}^{\text{f}}(\tau )=\left\{ 
\begin{array}{ll}
X_{\text{d}}(\tau ) & \text{for }\tau \leq \tau _{\!\partial \widehat{D}%
_{2}^{1}} \\ 
X_{\text{d}}(\tau _{\!\partial \widehat{D}_{2}^{1}}) & \text{for }\tau >\tau
_{\!\partial \widehat{D}_{2}^{1}}
\end{array}
\right. \;. 
\]
In fig. 1, the feasibility of the developed sampling technique is indicated
by graphical comparison of the numerically sampled and rigorous transition
probability densities $\rho _{\text{f}}^{\text{b}}$ for a given initial $%
\overline{x}^{\,\prime }$ denoted by the crosses.

\subsection{Free Fermion and Free Boson Diffusion on $\widehat{D}_2^3$}

We now formulate the diffusion process of two free indistinguishable
particles in three dimensions on the reduced state space $\widehat{D}%
_{2}^{3} $ by extending the symmetry analysis performed above. Following the
basic principles outlined in \cite{BDL96}, the free diffusion of
indistinguishable particles can be traced back to one-dimensional free
diffusion processes. In that framework, the transition probabilities $X_{%
\text{f,b}}^{\text{f,b}}(\tau )$ represent the elementary subprocesses
combined to constitute the overall free diffusion process in three
dimensions. For free identical fermions in three dimensions, the density
matrix can be expressed as 
\[
\rho (\overline{r},\tau ;\overline{r}^{\,\prime })=\frac{1}{8}\left[ 
\begin{array}{l}
\left[ \rho _{\text{f}}^{\text{f}}(\overline{x},\tau ;\overline{x}^{\,\prime
})\,+\rho _{\text{f}}^{\text{b}}(\overline{x},\tau ;\overline{x}^{\,\prime })%
\right] \left[ \rho _{\text{b}}^{\text{f}}(\overline{y},\tau ;\overline{y}%
^{\,\prime })\,+\rho _{\text{b}}^{\text{b}}(\overline{y},\tau ;\overline{y}%
^{\,\prime })\right] \left[ \rho _{\text{b}}^{\text{f}}(\overline{z},\tau ;%
\overline{z}^{\,\prime })\,+\rho _{\text{b}}^{\text{b}}(\overline{z},\tau ;%
\overline{z}^{\,\prime })\right] \\ 
+\left[ \rho _{\text{b}}^{\text{f}}+\rho _{\text{b}}^{\text{b}}\right] \left[
\rho _{\text{f}}^{\text{f}}+\,\rho _{\text{f}}^{\text{b}}\right] \left[ \rho
_{\text{b}}^{\text{f}}\,+\rho _{\text{b}}^{\text{b}}\right] +\left[ \rho _{%
\text{b}}^{\text{f}}\,+\rho _{\text{b}}^{\text{b}}\right] \left[ \rho _{%
\text{b}}^{\text{f}}+\,\rho _{\text{b}}^{\text{b}}\right] \left[ \rho _{%
\text{f}}^{\text{f}}\,+\rho _{\text{f}}^{\text{b}}\right] +\left[ \rho _{%
\text{f}}^{\text{f}}+\rho _{\text{f}}^{\text{b}}\right] \left[ \rho _{\text{f%
}}^{\text{f}}+\,\rho _{\text{f}}^{\text{b}}\right] \left[ \rho _{\text{f}}^{%
\text{f}}\,+\rho _{\text{f}}^{\text{b}}\right]
\end{array}
\right] 
\]
\noindent%
%
whereas for the corresponding boson system the density matrix evaluates to 
\[
\rho (\overline{r},\tau ;\overline{r}^{\,\prime })=\frac{1}{8}\left[ 
\begin{array}{l}
\left[ \rho _{\text{b}}^{\text{f}}(\overline{x},\tau ;\overline{x}^{\,\prime
})\,+\rho _{\text{b}}^{\text{b}}(\overline{x},\tau ;\overline{x}^{\,\prime })%
\right] \left[ \rho _{\text{b}}^{\text{f}}(\overline{y},\tau ;\overline{y}%
^{\,\prime })\,+\rho _{\text{b}}^{\text{b}}(\overline{y},\tau ;\overline{y}%
^{\,\prime })\right] \left[ \rho _{\text{b}}^{\text{f}}(\overline{z},\tau ;%
\overline{z}^{\,\prime })\,+\rho _{\text{b}}^{\text{b}}(\overline{z},\tau ;%
\overline{z}^{\,\prime })\right] \\ 
+\left[ \rho _{\text{f}}^{\text{f}}+\rho _{\text{f}}^{\text{b}}\right] \left[
\rho _{\text{f}}^{\text{f}}+\,\rho _{\text{f}}^{\text{b}}\right] \left[ \rho
_{\text{b}}^{\text{f}}\,+\rho _{\text{b}}^{\text{b}}\right] +\left[ \rho _{%
\text{b}}^{\text{f}}+\rho _{\text{b}}^{\text{b}}\right] \left[ \rho _{\text{f%
}}^{\text{f}}+\,\rho _{\text{f}}^{\text{b}}\right] \left[ \rho _{\text{f}}^{%
\text{f}}\,+\rho _{\text{f}}^{\text{b}}\right] +\left[ \rho _{\text{f}}^{%
\text{f}}\,+\rho _{\text{f}}^{\text{b}}\right] \left[ \rho _{\text{b}}^{%
\text{f}}+\,\rho _{\text{b}}^{\text{b}}\right] \left[ \rho _{\text{f}}^{%
\text{f}}\,+\rho _{\text{f}}^{\text{b}}\right]
\end{array}
\right] 
\]
In this representation, the density matrix for identical particles contains
32 terms. Since each term factorizes into a product of three one-dimensional
orthogonal density matrices, each of the 32 elementary three-dimensional
density matrices satisfies the Markov properties and generates a Wiener
subprocess on the reduced state space $\widehat{D}_{2}^{3}$. For instance,
the subprocess $\widehat{X}_{\text{fbb}}^{\text{fbf}}(\tau )$ 
\[
\widehat{X}_{\text{fbb}}^{\text{fbf}}(\tau )=\widehat{X}_{\text{f}}^{\text{f}%
}(\tau )\otimes \widehat{Y}_{\text{b}}^{\text{b}}(\tau )\otimes \widehat{Z}_{%
\text{b}}^{\text{f}}(\tau )\;, 
\]
\noindent can be sampled as the diffusion process of two free
distinguishable particles restricted to the reduced state space by
reflection at the boundaries $y_{1}=y_{2}$ and $z_{1}=z_{2}$ and absorption
at $x_{1}=x_{2}$, $x_{1}=-x_{2}$, $y_{1}=-y_{2}$ and $z_{1}=-z_{2}$. The
combination of all the 64 Wiener subprocesses gives rise to the free
diffusion process of two indistinguishable particles in three dimensions
defined on $\widehat{D}_{2}^{3}$ . The symmetry properties of the elementary
processes allow furthermore to unfold the resulting probability density into
the configuration space ${\Bbb R}^{6}$.

\subsection{Evolution Equations and Decomposition of the Potential}

The derivation of the evolution equation for the propagator proceeds along
the same lines as for the two-dimensional case, as discussed in Sec. II.A.2.
Without going into detail, we only mention the result 
\[
\frac{\partial }{\partial \tau }\,K_{I}[\overline{r}_{f},l{\bf ,}l^{\prime }%
{\bf ,}\tau ;\overline{r}_{i},\hat{l}\,,\hat{l}^{\prime },0]=\frac{\hbar }{2m%
}\,\nabla _{\overline{r}{\bf \,}}^{2}\,K_{I}[\overline{r}_{f},l{\bf ,}l{\bf %
^{\prime },}\tau ;\overline{r}_{i},\hat{l},\hat{l}^{\prime },0]-\frac{1}{
\hbar }\sum_{\overline{l},\overline{l}^{\prime }}V_{l\overline{l}%
}^{l^{\prime }\overline{l}^{\prime }}(\overline{r}_{f}){\bf \;}K_{I}[%
\overline{r}_{f},\overline{l}{\bf ,}\overline{l}^{\prime }{\bf ,}\tau ;%
\overline{r}_{i},\hat{l},\hat{l}^{\prime },0]\,, 
\]
where 
\[
K_{I}[\overline{r}_{f},l{\bf ,}l^{\prime }{\bf ,}\tau ;\overline{r}_{i},\hat{
l}\,,\hat{l}^{\prime },0]=\left( \frac{1}{2!}\right) ^{6}\sum_{\widehat{P}%
_{x}}\xi _{l^{\prime }}^{\widehat{P}_{x}}\sum_{\widehat{P}_{y}}\xi
_{l^{\prime }}^{\widehat{P}_{y}}\sum_{\widehat{P}_{z}}\xi _{l^{\prime }}^{%
\widehat{P}_{z}}\sum_{P_{x}}\xi _{l}^{P_{x}}\sum_{P_{y}}\xi
_{l}^{P_{y}}\sum_{P_{z}}\xi _{l}^{P_{z}}K_{D}[\widehat{P}_{x}\widehat{P}_{y}%
\widehat{P}_{z}P_{x}P_{y}P_{z}\,\overline{r}_{f},\tau ;\overline{r}_{i},\hat{
l},0]\quad . 
\]
The components of the potential are given by 
\[
V_{l\,\overline{l}}^{l^{\prime }\overline{l}^{\prime }}(\overline{r}_{f})%
{\bf \,}=\left( \frac{1}{2!}\right) ^{6}\sum_{\widehat{S}_{x}}\sum_{\widehat{
S}_{y}}\sum_{\widehat{S}_{z}}\sum_{S_{x}}\sum_{S_{y}}\sum_{S_{z}}\xi _{l}^{%
\widehat{S}_{x}}\xi _{l}^{\widehat{S}_{y}}\xi _{l}^{\widehat{S}_{z}}\xi
_{l^{\prime }}^{S_{x}}\xi _{l^{\prime }}^{S_{y}}\xi _{l^{\prime
}}^{S_{z}}\,V(\widehat{S}_{x}\widehat{S}_{y}\widehat{S}_{z}S_{x}S_{y}S_{z}\,%
\overline{r}_{f})\,\xi _{\bar{l}}^{S_{x}}\xi _{\bar{l}}^{S_{y}}\xi _{\bar{l}%
}^{S_{z}}\xi _{\bar{l}^{\prime }}^{\widehat{S}_{x}}\xi _{\bar{l}^{\prime }}^{%
\widehat{S}_{y}}\xi _{\bar{l}^{\prime }}^{\widehat{S}_{z}}\quad , 
\]
it should be noted that if the range of the label $l$ used to denoted the
statistics is extented in such a way that it also indicates the boundary
conditions on the reduced domain then the evolution equation for the
propagator derived here takes the same form as the equation (\ref{evcmdf}).

\section{An Algorithmic approach on Models}

In the previous section, we introduced a novel representation of the
Euclidean-time propagator of systems containing two indistinguishable
particles. This representation allows to sample the propagator
sign-problem-free. We now develop a numerically feasible sign-problem-free
implementation scheme. Its efficiency in numerical practice is analyzed in
applications to model systems. In all the cases, we use atomic units $\hbar
=m=e=1$. Energy and Euclidean time are measured in Hartrees (H) and H$^{-1}$.

\subsection{General Structure}

The key elements of the many-body diffusion algorithm (MBDA) for the
symmetry problems considered here are (i) an appropriately adapted free
diffusion step, (ii) a routine to model subprocess transitions, and (iii) a
branching and killing procedure. Whilst some elements of the algorithm
resemble Diffusion Monte Carlo, we stress that the underlying principle of
interdependent subprocess evolution is conceptually new. We are also aware
that the efficiency of the presented algorithm can be drastically enhanced
by importance sampling methods. Although we can show that this expectation
holds true, we do not dwell on these refinements, for major emphasis is on
the realization of symmetry arguments. In the limit of infinitely long
evolution $\tau \rightarrow \infty $, the stepwise application of the
Euclidean time propagator on a given sample leads in principle to the
system's lowest eigenstate compatible with the symmetry restrictions made.
In numerical practice, this evolution occurs in discrete Euclidean time
steps $\Delta \tau $. Since the assumption of such time steps is only exact
for $\Delta \tau \rightarrow 0$, the algorithm in principle suffers from a
systematic time-step error. The latter, however, could be controlled in the
considered applications. The main advantages of the developed approach are
the simplicity and convergence properties. In particular, it allows for the
transparent implementation of reflecting and absorbing boundaries to model
the subprocesses derived above. Let us discuss the algorithm in some more
detail. Having generated an initial population of walkers located at
positions $(\bar{r}_i^{\,\prime })_i $ in the appropriate state space, each
valid walker makes a free distinguishable particle move to $(\tilde{r}%
_f^{\,\prime })_i$. This is achieved by the construction of Gaussian
deviates with variance $\sqrt{\Delta \tau }$ and mean $\bar{r}_i^{\,\prime }$%
. If a walker hits the state-space boundary by free distinguishable particle
diffusion it is either reflected or absorbed depending on its characteristic
boundary conditions \cite{LBDL97}. In numerical practice, the use of
non-zero Euclidean time steps $\Delta \tau $ prevents the proper
construction of reflecting and absorbing boundaries. In the mathematical
literature, free diffusion in the presence of an absorbing boundary is
realized by the so-called Skohorod construction \cite{SKOHO}. The latter
reflects walker trajectories without changing the ``momentum'' of the
walker. This is in contrast to the reflection experienced by a classical
object at a hard wall. In the numerical procedures, the final reflected
position $\bar{r}_f^{\,\prime }$ is obtained by reflection of the
corresponding $\tilde{r}_f^{\,\prime }$. The systematic error for
inappropriate reflection vanishes for $\Delta \tau \rightarrow 0$; for the
utilized time intervals $\Delta \tau $ no significant error were observed.
The efficient implementation of fermionlike diffusion, in contrast, calls
for numerical crossing-recrossing corrections. For two identical fermions $%
j,k$ on a line the corrections for an absorbing boundary $x_j=x_k$ have been
discussed in \cite{LBDL97}. Based on the principle of images, we showed that
an apparently valid walker move from $(x_j^{\prime },x_k^{\prime })$ to $%
(x_j,x_k)$, both points inside the state space, has to be rejected with the
probability $\exp (-(x_j-x_k)(x_j^{\prime }-x_k^{\prime })/\Delta \tau )$.
The symmetry properties of the potentials discussed in this work require the
generation of interdependent subprocesses. In succession to the adapted free
diffusion step, a walker associated with the boundary condition of the
subprocess $l$ may change to the boundary condition of a subprocess $%
l^{\prime }$ with the probability $p_{l\rightarrow l^{\prime }}(\bar{r}_f)$.
Its actual structure is determined by the physical situation, e.g. the
symmetry properties of both the potential and the eigenstate under
consideration. The local dependence of the probabilities $p_{l\rightarrow
l^{\prime }}(\bar{r}_f)$ governs the relevance of subprocess transitions at
a given position $\bar{r}_f$. The structure of $p_{l\rightarrow l^{\prime }}(%
\bar{r}_f)$ is fairly flexible as long as it harmonizes with the killing and
branching procedure. For instance, $p_{l\rightarrow l^{\prime }}(\bar{r}_f)$
could be decoupled from the different kinds of boundary conditions. Then
each subprocess transition occurs with equal probability $1/n_l$, where $n_l$
denotes the total number of subprocesses. Accordingly, both the
normalization rates and the probability for branching and killing become
dependent on the subprocess transition performed on the walker. In what
follows, we assume subprocess transition probabilities adapted to the
branching and killing procedure due to the potential. The third step,
branching and killing, realizes the presence of the potential by the
implementation of the exponential path-integral weights $w(\bar{r}^{\,\prime
},\bar{r})=\left. \exp \left( -\int\limits_0^{\Delta \tau }V\left( \overline{
R}(\varsigma )\right) d\varsigma \right) \right| _{\overline{R}(0)=\bar{r}%
^{\,\prime }}^{\overline{R}(\Delta \tau )=\bar{r}}$, where $\overline{R}%
(\varsigma )$ denotes the paths generated by the adapted free diffusion
process. The principal lack of continuous trajectories in discrete time-step
evolution procedures induces the necessity for feasible approximations to $w(%
\bar{r}^{\,\prime },\bar{r})$. For sufficiently smooth potentials, the
Suzuki-Trotter formula \cite{SUZUK} offers a reliable alternative. Then only
the values of the potential at the initial and final points of a move are
taken into account: $w(\bar{r}^{\,\prime },\bar{r})=\exp \{-(V(\bar{r}%
^{\,\prime })+V(\bar{r}))\Delta \tau /2\}$. Especially for singular
potentials, improved approximations (for a detailed discussion see e.g. \cite
{CEP95} and references therein) are recommended. Here, we choose the
semiclassical approximation suggested in \cite{MAK88}: $w(\bar{r}^{\,\prime
},\bar{r})=\exp \{-\int_0^1d\xi \,V(\bar{r}^{\,\prime }+\xi (\bar{r}-\bar{r}%
^{\,\prime }))\}$. A branching and killing technique serves to efficiently
adapt the walkers to the weights $w(\bar{r}^{\,\prime },\bar{r})$ by an
acceptance/rejection procedure. To conserve the size of the walker
population to within statistical fluctuations, an appropriate reference
energy $E_{\text{ref}}$ has been supplied to yield improved weights $w(\bar{r%
}^{\,\prime },\bar{r})\,e^{E_{\text{ref}}\Delta \tau }$. In \cite{LBDL97},
we derived an estimator for the ground-state energy $E_0$. The key elements
of this estimator are the potential average $\left\langle V\right\rangle _D$
over the state space $D$ and a surface term $\left\langle j\right\rangle
_{\partial D}$ triggered by the non-zero gradient of the fermionlike
subdensities at absorbing boundaries. The estimate is easily generalized to
an ensemble of interdependent subdensities $\Psi (\overline{r},l{\bf ,}\tau
) $%
\[
E_0\,=\,\lim_{\tau \to \infty }E_\tau \;=\;\lim_{\tau \to \infty }\sum_l\;
\left[ -\frac{\frac{\hbar ^2}{2m}\int_{\partial D_2}d\overline{r}\;\overline{%
\nabla }_{\overline{r}}\Psi (\overline{r},l{\bf ,}\tau )}{ \int_{D_2}d%
\overline{r}\;\Psi (\overline{r},l{\bf ,}\tau )}+\frac{\int_{D_2}d\overline{r%
}\;\Psi (\overline{r},l{\bf ,}\tau )\,V(\overline{r})}{\int_{D_2}d\overline{r%
}\;\Psi (\overline{r},l{\bf ,}\tau )}\right] \,=\,\sum_l\left[ \langle
\,j\,\rangle _{\partial D_2}^l+\langle \,V\,\rangle _{D_2}^l\right] \ \ \ . 
\]
Alternatively, the growth estimates 
\[
E_0\,=\,\lim_{\tau \to \infty }E_\tau =-\lim_{\tau \to \infty }\frac 1\tau
\ln \int d\overline{r}_f\int d\overline{r}_i\sum_lK[\overline{r}_f,l{\bf ,}%
\tau ;\overline{r}_i,0]\,\ \ , 
\]
\noindent may serve to predict the ground-state energy. Based on this
general algorithmic structure, the ground-state wave function of several
model systems is investigated below. The different types of potentials make
a specific analysis of the appropriate state space necessary.

\subsection{Double well and Mexican Hat}

To start with, the symmetry principles underlying the many-body diffusion
algorithm are illustrated for a quantum-mechanical particle in a double-well
or a Mexican Hat potential. For these models, the utilization of
(anti)symmetric subprocesses is shown to provide a suitable technique for
the numerical generation of the real ground-state wave function. Our study
aims at revealing the role of symmetry arguments to sample a probability
density on an appropriate part of the configuration space. Feasibility of
the symmetry concept has been tested independently with standard Monte Carlo
techniques. Consider a single particle in a one-dimensional double-well
potential, described by the Hamiltonian 
\begin{equation}
H=-\frac 12\frac{\partial ^2}{\partial x^2}+V(x)\quad \text{with}\quad V(x)=%
\frac{(x^2-a)^2-a^2}4+b\,x^3\;,  \label{double-well}
\end{equation}
\noindent where $a$ and $b$ represent parameters. Fig. 2 depicts the
potential values for two different parameter choices: $a=2,b=0$ and $%
a=2,b=0.25$ in the following referred to as symmetric and asymmetric
potential. According to the rules outlined above, the Euclidean time
propagator $K(x,\tau ;x^{\,\prime })$ associated with the Hamiltonian (\ref
{double-well}) can be formulated as a sum of four subpropagators $K(x,\tau
,\jmath ;x^{\,\prime },\tilde{\jmath})$ with $\jmath ,\tilde{\jmath}\in
\{0,1\}$ denoting symmetric resp. antisymmetric behavior under the
transformation $\overline{P}_x:x\rightarrow -x$. This means that $K(x,\jmath
,\tau ;x^{\,\prime },0)$ and $K(x,\jmath ,\tau ;x^{\,\prime },1)$ are
initially symmetric and antisymmetric with respect to the boundary $x=0$
which separates the two half-spaces $D^{-}=\{x|x\leq 0\}$ and $%
D^{+}=\{x|x\geq 0\}$. The (anti)symmetrized potentials $V_{\text{s}}(x)$ and 
$V_{\text{a}}(x)$ are derived as 
\[
V_{\text{s}}(x)\equiv V_{\jmath \jmath }(x)=\frac{V(x)+V(-x)}2=\frac{
x^2(x^2-2a)}4\;;\;V_{\text{a}}(x)\equiv V_{\jmath \jmath ^{\prime }}(x)=%
\frac{V(x)-V(-x)}2=b\,x^3\;,\jmath ,\jmath ^{\prime }\in \{0,1\},\jmath \neq
\jmath ^{\prime } 
\]
The matrix $W$ (cfr. (\ref{ndrate})) is calculated as $W_{\jmath \jmath
^{\prime }}=-\,V_{\text{a}}\,(1-\delta _{\jmath ,\jmath ^{\prime }})$ with
the total rate represented by $W_{\jmath }=-\,V_{\text{a}}$. The
Euclidean-time Schr\"{o}dinger equation [cfr. (\ref{evcmdf})] can thus be
written 
\begin{equation}
\frac \partial {\partial \tau }\,K[x,\jmath {\bf ,}\tau ;x^{\prime },\hat{
\jmath},0]=\frac 12\,\frac{\partial ^2}{\partial x^2}\,K[x,\jmath {\bf ,}%
\tau ;x^{\prime },\hat{\jmath}\,,0]-V_{\text{s}}(x)\,K_I[x,\jmath {\bf ,}%
\tau ;x^{\prime },\hat{\jmath},0]-\sum_{\jmath ^{\prime }=0}^1V_{\text{a}%
}(x)\,K[x,\jmath ^{\prime }{\bf ,}\tau ;x^{\prime },\hat{\jmath}%
\,,0]\,(1-\delta _{\jmath ,\jmath ^{\prime }})\,.  \label{evol}
\end{equation}
Eq. (\ref{evol}) allows for a change of the boundary conditions of the
propagator $K[x,\jmath {\bf ,}\tau ;x^{\prime },\hat{\jmath},0]$ during its
evolution in Euclidean time. Written as a Feynman-Kac functional, the
propagator $K[x,\jmath {\bf ,}\tau ;x^{\prime },\hat{\jmath},0]$ can be
interpreted as a sum of two interdependent distinguishable-particle
diffusion subprocesses. The numerical realization of the subprocesses $%
\jmath $ occurs by means of ensembles $(x_i^{\jmath })_{i=1,n_{\jmath }}$ of 
$n_{\jmath }$ walkers. Each of the $n_{\jmath }$ walkers satisfies the
boundary conditions for the subprocess $\jmath $ and during the evolution a
walker may change its subprocess $\tilde{\jmath}$ to $\jmath $ according to
the normalized transition amplitudes $p_{\jmath \rightarrow \jmath ^{\prime
}}(\bar{r}_f)$. To formulate a probability-based evolution scheme it is
therefore required that $p_{\jmath \rightarrow \jmath ^{\prime }}(\bar{r}_f)$
remains positive. Since 
\[
\sum_{\bar{\jmath}=0}^1\exp \left( -\,\tau \,V_{\jmath ^{\prime }\bar{\jmath}
}(\bar{r}_f)\right) =2\exp \left( -\,\tau \,\frac{(x^2-a)^2-a^2}4\right) %
\left[ \cosh \left( -\,\tau \,b\,x^3\right) \delta _{\jmath ^{\prime
},\jmath }+\sinh \left( -\,\tau \,b\,x^3\right) (1-\delta _{\jmath ^{\prime
},\jmath })\right] , 
\]
\noindent the half-space $D^{-}$ represents the appropriate state space for
probabilistic sampling. Walker transitions to $x>0$ are prevented by the
construction of a reflecting (for $\jmath {\bf =}0$) or an absorbing (for $%
\jmath {\bf =}1$) boundary. The major algorithmic steps for the numerical
evolution procedure are enlisted as follows:

\begin{enumerate}
\item  Generate two distinct walker sets $(x_{i}^{\jmath })_{i=1,n_{\jmath }}
$ , $\jmath {\bf =}0,1$. The initial non-zero walker populations $n_{\jmath }
$ are chosen randomly.

\item  Calculate for each walker the subprocess-transition probabilities $p_{%
\text{a}}(x)=\,\left[ 1-\exp \left( -2\Delta \tau bx^{3}\right) \right] /2$
and apply them by comparison with uniform pseudo-random numbers $\eta \in
\lbrack 0,1]$: if $\eta \leq p_{\text{f}}(x)$ change the boundary condition
associated to the walker, else leave the condition unchanged.

\item  Make an evolution step, as indicated in Sec. A.

\item  Return to step 1. until the density has equilibrated; after a certain
amount of repetitions, calculate the potential average and the walker
outflow to get an estimate for the ground-state energy.

\item  To obtain an image of the sampled ground-state wave function, record
the walker positions in small discrete bins.
\end{enumerate}

The results obtained for the ground-state energy $E_0$ of the double-well
system are shown in table 1. To test the algorithm, the estimates $E_0^{%
\text{MBDA}}$ have been compared with the numerical outcome $E_0^{\text{conf}
}$ of a program with evolution on the full configuration space. For both the
symmetric and the asymmetric potential, the results agree within the
estimated standard deviation $\sigma $. It is illuminating to examine the
population sizes of the two subprocesses $\jmath $. For the symmetric
potential the probability $p_{\text{a}}(x)$ is zero and independent
subprocesses are simulated. Due to the higher eigenenergy related to the
fermion subprocess $K[x,0{\bf ,}\tau ;x^{\prime },\hat{\jmath},0]$, the
antisymmetric process rapidly fades out and the ground state wave function
is completely described by the symmetric process. For the asymmetric
potential in contrast, the antisymmetric subprocess is of crucial
importance, since the asymmetric ground-state wave function cannot be simply
generated by a symmetric process. The faster decay of the antisymmetric
density is counteracted by a net walker transition from the symmetric to the
antisymmetric population. In equilibrium, almost half of the population
belongs to the antisymmetric density.

\begin{center}
\hspace{0.2in}\hspace{-1in} 
\begin{tabular}{c||c|c|c|c|c|c||}
& $E_0^{\text{conf}}$ & $\sigma ^{\text{conf}}$ & $E_0^{\text{MBDA}}$ & $%
\sigma ^{\text{MBDA}}$ & $\sum \Psi _{\text{s}}/\sum \Psi $ & $\sum \Psi _{%
\text{a}}/\sum \Psi $ \\ \hline\hline
\multicolumn{1}{l||}{$b=0$} & -0.2996 & 0.0006 & -0.2999 & 0.0006 & 100 \% & 
0 \% \\ 
\multicolumn{1}{l||}{$b=0.25$} & -1.0251 & 0.0012 & -1.0251 & 0.0011 & 53.7
\% & 46.3 \% \\ \hline\hline
\end{tabular}
\quad {\bf Table 1} \medskip
\end{center}

This example for a particle in one dimension can be transparently
generalized to arbitrary spatial dimensions. Consider, e.g., a particle in
two dimensions exposed to the Mexican-Hat potential 
\begin{equation}
H=-\frac 12\frac{\partial ^2}{\partial x^2}-\frac 12\frac{\partial ^2}{%
\partial y^2}+V(x,y)\quad \text{with}\quad V(x,y)=\frac{(x^2+y^2-a)^2-a^2}4
+b\,x^3\;.  \label{MexiHat}
\end{equation}
The two parameter sets $a=2,b=0$ and $a=2,b=0.25$ were chosen as in one
dimension to represent a symmetric and an asymmetric potential, as depicted
in fig. 2. Consider the two operators $\overline{P}_x:x\rightarrow -x$ and $%
\overline{P}_y:y\rightarrow -y$ and introduce the two boundaries $x=0$ and $%
y=0$. The construction of reflecting and absorbing boundary conditions at
these two boundaries calls for the definition of 16 Euclidean-time
propagators $K(x,\tau ,\jmath ;x^{\,\prime },\hat{\jmath})$ and four
elementary states $\Psi _{\text{ss}}$, $\Psi _{\text{sa}}$, $\Psi _{\text{as}
}$ and $\Psi _{\text{aa}}$ with their indices characterizing their
properties under transformation with $\overline{P}_x$ and $\overline{P}_y$,
respectively. By imposing reflecting and absorbing boundary conditions the
configuration space ${\Bbb R}^2$ is split into four subspaces with positive
or negative $x$- or $y$-coordinates. Although the system's evolution can be
folded to any of these four subspaces, the analysis of the
subprocess-transition amplitudes for the Mexican-Hat potential reveals that
only two subspaces are associated with positive definite sampling. A sign
analysis of the four subprocess-transition amplitudes $p_{\text{ss}}(\bar{r}
_f)$, $p_{\text{sa}}(\bar{r}_f)$, $p_{\text{as}}(\bar{r}_f)$ and $p_{\text{aa%
}}(\bar{r}_f)$ 
\[
p_{\text{as}}(x)=\,\frac{1-\exp \left( -2\Delta \tau bx^3\right) }2\;,\;p_{%
\text{ss}}(x)=\,\frac{1+\exp \left( -2\Delta \tau bx^3\right) }2\;,\;p_{%
\text{sa}}(x)=0\;,\;p_{\text{aa}}(x)=0\;, 
\]
\noindent indicates that they all may be interpreted as probabilities if the
evolution is confined to either the subspace $D_{-}^{+}\equiv \{x\leq
0,y\geq 0\}$ or $D_{-}^{-}\equiv \{x\leq 0,y\leq 0\}$. The fact that $\;p_{%
\text{sa}}(x)=0$ and $p_{\text{aa}}(x)=0$ induces two independent sets of
subprocesses, namely [$K(x,\tau ,$ss$;x^{\,\prime },\hat{\jmath})$, $%
K(x,\tau ,$as$;x^{\,\prime },\hat{\jmath})$] and [$K(x,\tau ,$sa$%
;x^{\,\prime },\hat{\jmath})$, $K(x,\tau ,$aa$;x^{\,\prime },\hat{\jmath})$%
]. Only those subprocess transitions are allowed which conserve the parity
under $\overline{P}_y$. The formulation of the respective MBDA yields
results for the ground-state energy which are in good agreement with the
results achieved without the introduction of boundaries (see table 2). Since
the characteristics of the model (\ref{MexiHat}) allow for a
distinguishable-particle treatment, the boundary-free algorithm reliably
simulates the properties under investigation. The comparison of the
estimated ground-state energies demonstrates the efficiency of the MBDA.
Both energies $E_0^{\text{conf}}$ and $E_0^{\text{MBDA}}$ have been obtained
with comparable numerical effort and standard deviations. The last four
columns of table 2 show the relevance of the four states $\Psi _{\text{ss}}$
, $\Psi _{\text{sa}}$, $\Psi _{\text{as}}$ and $\Psi _{\text{aa}}$ with as
measure of their relative population sizes. Since $p_{\text{sa}}(x)=0$ and $%
p_{\text{aa}}(x)=0$, the states $\Psi _{\text{sa}}$ and $\Psi _{\text{aa}}$
fade out. In addition, for the symmetric potential $p_{\text{as}}(x)=0$.
This means that also $\Psi _{\text{as}}$ is irrelevant for the sampling of
the ground-state wave function. The remaining state $\Psi _{\text{ss}}$ is
symmetric under both $\overline{P}_x$ and $\overline{P}_y$. Clearly,
evolution in the presence of the asymmetric potential cannot be simulated
with a symmetric state only; the algorithm ensures the occurrence of the $%
\Psi _{\text{as}}$ state by a steady transformation of walkers associated
with the subprocess $K(x,\tau ,$ss$;x^{\,\prime },\hat{\jmath})$ to walkers
propagating according to $K(x,\tau ,$as$;x^{\,\prime },\hat{\jmath})$.

\begin{center}
\bigskip 
\begin{tabular}{c||c|c|c|c|c|c|c|c||}
& $E_0^{\text{conf}}$ & $\sigma ^{\text{conf}}$ & $E_0^{\text{MBDA}}$ & $%
\sigma ^{\text{MBDA}}$ & $\sum \Psi _{\text{ss}}/\sum \Psi $ & $\sum \Psi _{%
\text{as}}/\sum \Psi $ & $\sum \Psi _{\text{sa}}/\sum \Psi $ & $\sum \Psi _{%
\text{aa}}/\sum \Psi $ \\ \hline\hline
\multicolumn{1}{l||}{$b=0$} & -0.19865 & 0.00042 & -0.19860 & 0.00037 & 100
\% & 0 \% & 0 \% & 0 \% \\ 
\multicolumn{1}{l||}{$b=0.25$} & -0.56129 & 0.00054 & -0.56115 & 0.00050 & 
60.5 \% & 39.5 \% & 0 \% & 0 \% \\ \hline\hline
\end{tabular}
\quad {\bf Table 2}
\end{center}

\subsection{Anisotropic Harmonic Fermions}

A system of two non-interacting anisotropic identical fermion oscillators in
three dimensions serves as a transparent testing ground for
indistinguishable particles. We considered a system with the Hamiltonian 
\begin{equation}
H=\frac{\overline{p}^{\,2}}2+\frac{\Omega _x^2}2\left( x_1^2+x_2^2\right) +%
\frac{\Omega _y^2}2\left( y_1^2+y_2^2\right) +\frac{\Omega _z^2}2\left(
z_1^2+z_2^2\right) \;,  \label{mode}
\end{equation}
with ground-state energy 
\begin{equation}
E_0=\Omega _x+\Omega _y+\Omega _z+\text{Min[}\Omega _x,\Omega _y,\Omega _z%
\text{]\ .}  \label{ex1}
\end{equation}
We now consider a rotated reference frame as the result of from the rotation
by the two Euler angles $\phi $ and $\theta $ \cite{GOLD} 
\[
\overline{R}=\left[ 
\begin{array}{ccc}
\cos \phi & \sin \phi & 0 \\ 
-\sin \phi \cos \theta & \cos \phi \cos \theta & 0 \\ 
\sin \phi \sin \theta & -\cos \phi \sin \theta & \cos \theta
\end{array}
\right] \;\overline{r}\ . 
\]
The objectives are clear: the symmetry of the potential under the
interchange of Cartesian particle coordinates is avoided without losing the
advantage of comparison with a rigorous analytical solution. A proper
implementation of the propagator with the MBDA requires positive definite
subprocess transition amplitudes 
\begin{equation}
p_{l\overline{l}}(\bar{r}_f)=\frac{\sum_{S_x}\sum_{S_y}\sum_{S_z}\xi
_l^{S_x}\xi _l^{S_y}\xi _l^{S_z}\exp \left( -\tau V(S_xS_yS_z\,\overline{r}%
_f)\right) \xi _{\overline{l}}^{S_x}\xi _{\overline{l}}^{S_y}\xi _{\overline{%
l}}^{S_z}}{\sum_{\overline{l}}\sum_{S_x}\sum_{S_y}\sum_{S_z}\xi _l^{S_x}\xi
_l^{S_y}\xi _l^{S_z}\exp \left( -\tau V(S_xS_yS_z\,\overline{r}_f)\right)
\xi _{\overline{l}}^{S_x}\xi _{\overline{l}}^{S_y}\xi _{\overline{l}}^{S_z}}%
\;,  \label{prob2}
\end{equation}
\noindent where we used the notation introduced in Sec. II. If the numerator
of eq. (\ref{prob2}) is positive definite, $p_{l\overline{l}}(\bar{r}_f)$
may be interpreted as the probability for a subprocess transition $\overline{
l}\rightarrow l$. Since the sign of $p_{l\overline{l}}(\bar{r}_f)$ is
determined by the parity of the propagator $K[\overline{r}_f,${\bf $l$}${\bf %
,}\Delta \tau ;\overline{r}_i,\hat{l}{\bf ,}0]$ and the boundary condition $%
\overline{l}$ under Cartesian-coordinate permutation, four different
subprocess-transition amplitudes occur. Their classification according to
their symmetry properties suggests the following notation: $p_{\text{bbb}}(%
\overline{r}_f)$, $p_{\text{ffb}}(\bar{r}_f)$, $p_{\text{fbf}}(\bar{r}_f)$
and $p_{\text{bff}}(\bar{r}_f)$. The probability $p_{\text{ffb}}(\bar{r}_f)$%
, e.g., stands for transitions between subprocesses with even parity in the $%
z$-coordinate and odd parity in the $x,y$-coordinates, for instance $K[%
\overline{r}_f,$fbb${\bf ,}\Delta \tau ;\overline{r}_i,$bfb${\bf ,}0]$.
While the amplitude $p_{\text{bbb}}(\vec{r})$ is always positive, positivity
of the three remaining amplitudes is imposed (in first order of the
Euclidean time step $\Delta \tau $) by the following conditions on the
domain and the parameters $\Omega _x$, $\Omega _y$, $\Omega _z$, $\phi $ and 
$\theta $: 
\begin{equation}
\left\{ 
\begin{array}{ll}
\sin \phi \cos \phi \left( \Omega _x^2-\Omega _y^2\cos ^2\theta +\Omega
_z^2(\cos ^2\theta -1)\right) \,(x_1-x_2)\,(y_1-y_2) & \geq 0 \\ 
\cos \phi \sin \theta \cos \theta \;\left( \Omega _y^2-\Omega _z^2\right)
\,(y_1-y_2)\,(z_1-z_2) & \geq 0 \\ 
-\sin \phi \sin \theta \cos \theta \;\left( \Omega _y^2-\Omega _z^2\right)
\,(x_1-x_2)\,(z_1-z_2) & \geq 0
\end{array}
\right\} \ .  \label{cond}
\end{equation}
For the present illustration, the evolution has been restricted to the state
space $\widehat{D}_3^2$ with a particular set of parameters: 
\begin{equation}
x_1\geq x_2\,,\,y_1\geq y_2\,,\,z_1\geq z_2\,,\,\Omega _x=4\,,\,\Omega
_y=3\,,\,\Omega _z=2\,,\,\phi =130{{}^{\circ }}\,,\,\theta =50{{}^{\circ }}\
.  \label{para}
\end{equation}
Employing samples of approximately 20,000 walkers and Euclidean time steps
of 0.001 H$^{-1}$, the MBDA leads to the numerical results visualized in
fig. 5. The ground-state energy estimates (see left-hand side of fig. 5)
oscillate in a small interval around their average of 10.9987$\pm $0.0056 H.
They estimate the rigorous ground-state energy of 11.0 H with a statistical
inaccuracy of about half a per mil. After numerical thermalization, the
population of each of the four subprocesses \{f,b,b\}, \{b,f,b\}, \{b,b,f\}
and \{f,f,f\} remains approximately constant (see right-hand side of fig.
5). In particular, contributions from the \{f,f,f\}-subprocess prove
indispensable to correctly generate the propagator. Our numerical
investigations manifested the occurrence of a net walker flow from the
subdensities $\Psi _{\text{fbb}},\Psi _{\text{bfb}},\Psi _{\text{bbf}}$ to $%
\Psi _{\text{fff}}$. Without the inclusion of subprocess interdependencies,
the subprocess $K[\overline{r}_f,$fff${\bf ,}\Delta \tau ;\overline{r}_i,%
\hat{l}{\bf ,}0]$ would fade out exponentially in evolution time.

\subsection{Application to Ortho Helium}

To illustrate the feasibility of the reduced state space concept, we also
used the MBDA to calculate the ground-state energy of ortho-helium. In the
triplet states of ortho-helium, the two electrons carry equal spin. The
occurrence of additional repulsive forces due to exchange distinctly
influences the energy spectrum, so that the simulation of the lowest
ortho-helium state requires the accurate inclusion of Fermi statistics. In
that context, the challenge of numerical implementation shows twofold.
First, the lowest energy level should be reliably predicted and second, the
converged energy estimates should be stable in time. In particular the
second aspect concerns the fermion sign problem: the MBDA is demonstrated to
exactly filter out possible interfering symmetric contributions. The general
evolution scheme on the reduced state space $\widehat{D}_{\,2}^{\,3}$ relies
on the construction of 32$^{2}$ interdependent subprocesses $K[\overline{r}%
_{f},l,l^{\prime }{\bf ,}\Delta \tau ;\overline{r}_{i},\hat{l},\hat{l}%
^{\prime }{\bf ,}0].$ Here, each of the $l$,$\hat{l}$ may be one of the four
boundary conditions fff,fbb,bfb or bbf, whereas $l^{\prime },\hat{l}^{\prime
}$ run over the eight conditions bbb,fbb,..,fff. Symmetry arguments allow to
simplify the computational procedure by the prior rejection of certain
classes of subprocesses. As the permutation operator $P$ has the same effect
on the ortho-helium Hamiltonian as the permutation operator $\widehat{P}$,
the ground-state wave function consists of substates with equal parity under 
$P$ and $\widehat{P}$. Consequently, the simulation of the ortho-helium
ground-state wave function on the reduced state space $\widehat{D}%
_{\,2}^{\,3}$ is accomplished by the construction of only 16 subprocesses
with $l=\hat{l},$ $l^{\prime }=\hat{l}^{\prime }$ specified as fff,fbb,bfb
or bbf. Although both - the general and the simplified - procedure in
principle lead to the exact densities, the restriction to four subprocesses
considerably improves the convergence of the algorithm. The next step in the
formulation of an evolution scheme for the ortho-helium ground state is the
consideration of the subprocess-transition amplitudes 
\begin{equation}
p_{l\overline{l}}^{l^{\prime }\overline{l}^{\prime }}(\bar{r}_{f})=\frac{%
\sum_{\widehat{S}_{x},\widehat{S}_{y},\widehat{S}_{z},\widehat{S}_{x},%
\widehat{S}_{y},\widehat{S}_{z}}\xi _{l}^{\widehat{S}_{x}}\xi _{l}^{\widehat{%
S}_{y}}\xi _{l}^{\widehat{S}_{z}}\xi _{l^{\prime }}^{S_{x}}\xi _{l^{\prime
}}^{S_{y}}\xi _{l^{\prime }}^{S_{z}}\,e^{-\tau V(\widehat{S}_{x}\widehat{S}%
_{y}\widehat{S}_{z}S_{x}S_{y}S_{z}\,\overline{r}_{f})}\,\xi _{\overline{l}%
}^{S_{x}}\xi _{\overline{l}}^{S_{y}}\xi _{\overline{l}}^{S_{z}}\xi _{%
\overline{l}^{\prime }}^{\widehat{S}_{x}}\xi _{\overline{l}^{\prime }}^{%
\widehat{S}_{y}}\xi _{\overline{l}^{\prime }}^{\widehat{S}_{z}}}{\sum_{%
\overline{l},\overline{l}^{\prime },\widehat{S}_{x},\widehat{S}_{y},\widehat{%
S}_{z},\widehat{S}_{x},\widehat{S}_{y},\widehat{S}_{z}}\xi _{l}^{\widehat{S}%
_{x}}\xi _{l}^{\widehat{S}_{y}}\xi _{l}^{\widehat{S}_{z}}\xi _{l^{\prime
}}^{S_{x}}\xi _{l^{\prime }}^{S_{y}}\xi _{l^{\prime }}^{S_{z}}\,e^{-\tau V(%
\widehat{S}_{x}\widehat{S}_{y}\widehat{S}_{z}S_{x}S_{y}S_{z}\,\overline{r}%
_{f})}\,\xi _{\overline{l}}^{S_{x}}\xi _{\overline{l}}^{S_{y}}\xi _{%
\overline{l}}^{S_{z}}\xi _{\overline{l}^{\prime }}^{\widehat{S}_{x}}\xi _{%
\overline{l}^{\prime }}^{\widehat{S}_{y}}\xi _{\overline{l}^{\prime }}^{%
\widehat{S}_{z}}}  \label{amplit}
\end{equation}
\noindent controlling the interplay between the four single subprocesses. In
order to derive the subspaces associated with positive sums of exponentials $%
e^{-\tau V(\widehat{S}_{x}\widehat{S}_{y}\widehat{S}_{z}S_{x}S_{y}S_{z}\,%
\overline{r}_{f})}$, we focus on the short-time limits $(1-\tau V(\widehat{S}%
_{x}\widehat{S}_{y}\widehat{S}_{z}S_{x}S_{y}S_{z}\,\overline{r}_{f}))$ of
the exponentials, which contain all the essential symmetry characteristics
necessary for the sign analysis of the amplitudes (\ref{amplit}).
Furthermore, it proves useful to separate the potential $V(\overline{r})$ in
two parts $V_{1}(\overline{r})$ and $V_{2}(\overline{r})$ with 
\[
V_{1}(\overline{r})=\frac{-2}{|\vec{r}_{1}|}+\frac{-2}{|\vec{r}_{2}|}\ \ \
,\ \ \ V_{2}(\overline{r})=\frac{1}{|\vec{r}_{1}-\vec{r}_{2}|}\ \ \ ,\ \ \ V(%
\overline{r})=V_{1}(\overline{r})+V_{2}(\overline{r})\ .
\]
To illustrate the need of a reduced state space, consider the corresponding
many-body diffusion process on the state space $D_{\,2}^{\,3}$. The
amplitudes $p_{\text{bbb}}(\overline{r})$, $p_{\text{ffb}}(\overline{r})$, $%
p_{\text{fbf}}(\overline{r})$ and $p_{\text{bff}}(\overline{r})$ control the
transitions between the 16 fermion subprocesses $K[\overline{r}_{f},l{\bf ,}%
\Delta \tau ;\overline{r}_{i},\hat{l}{\bf ,}0]$. A detailed investigation of
the amplitudes, however, discloses that some of them may become negative on
the state space $D_{\,2}^{\,3}$. For instance, considering $p_{\text{ffb}}(%
\overline{r})$, one derives 
\begin{eqnarray*}
&&p_{\text{ffb}}(\overline{r})\stackrel{\tau \rightarrow 0}{:\longrightarrow 
}e^{\tau V(\,\overline{r})}\!\sum%
\limits_{P_{x},P_{y},P_{z}}(-)^{P_{x}}(-)^{P_{y}}\left(
1-V_{1}(P_{x}P_{y}P_{z}\overline{r})\right)  \\
&\propto &\frac{4\tau }{\sqrt{\pi }}\int_{-\infty }^{\infty }dt\,\left(
e^{-x_{1}^{\,2}t^{2}}-e^{-x_{2}^{\,2}t^{2}}\right) \left(
e^{-y_{1}^{\,2}t^{2}}-e^{-y_{2}^{\,2}t^{2}}\right) \left(
e^{-z_{1}^{\,2}t^{2}}+e^{-z_{2}^{\,2}t^{2}}\right) 
\end{eqnarray*}
which is positive definite on a subdomain of the state space, namely for $%
|x_{1}\!|\geq \!|x_{2}|$ and $|y_{1}\!|\geq \!|y_{2}|$. Similar restrictions
follow from $p_{\text{fbf}}(\overline{r})$ and $p_{\text{bff}}(\overline{r})$%
, whereas $p_{\text{bbb}}(\overline{r})$ is a sum of only positive elements.
Strictly speaking, utilizing the state space as the domain for evolution,
the MBDA would incorrectly simulate the ortho-helium propagator. It is at
that point that the reduced state space $\widehat{D}_{\,2}^{\,3}$ as defined
by (\ref{reduced}) comes into play. By appropriate restriction of the
evolution to the reduced state space, the subprocess-transition amplitudes
all remain positive and their probabilistic implementation is justified. In
what follows, we take over the results from Sec. III and adapt the general
scheme to the ortho-helium system.\\[0.2cm]
Continuing with the investigation of the amplitudes $p(\overline{r})$, let
us first focus on the contributions of $V_{1}(\overline{r})$. In $V_{1}(%
\overline{r})$, the coordinates $x_{1},\ldots ,z_{2}$ are entangled such
that the transformations $P$ and $\hat{P}$ have the same effects; for our
purpose, no distinction has to be made between $P$ and $\hat{P}$. \newline
With regard to amplitudes with identical sub- and superindices per
coordinate, one obtains 
\begin{eqnarray*}
\left. p_{\text{ffb}}^{\text{ffb}}(\overline{r})\right| _{V_{1}} &=&e^{\tau
V(\,\overline{r})}\hspace{-0.2in}\sum\limits_{P_{x},P_{y},P_{z},\widehat{P}%
_{x}\widehat{P}_{y},\widehat{P}_{z}}\hspace{-0.2in}%
(-)^{P_{x}}(-)^{P_{y}}(-)^{\widehat{P}_{x}}(-)^{\widehat{P}_{y}}\exp
(P_{x}P_{y}\widehat{P}_{x}\widehat{P}_{y}\overline{r}) \\
&\stackrel{\tau \rightarrow 0}{:\longrightarrow }&\frac{\tau }{2\sqrt{\pi }}%
\,e^{\tau V_{1}(\,\overline{r})}\int_{-\infty }^{\infty }dt\,\left(
e^{-x_{1}^{\,2}t^{2}}-e^{-x_{2}^{\,2}t^{2}}\right) \left(
e^{-y_{1}^{\,2}t^{2}}-e^{-y_{2}^{\,2}t^{2}}\right) \left(
e^{-z_{1}^{\,2}t^{2}}+e^{-z_{2}^{\,2}t^{2}}\right) \text{.}
\end{eqnarray*}
The remaining 28 amplitudes have in common that for at least one coordinate
the permutations $P$ and $\widehat{P}$ induce different parity. Each of
these subprocess-transition amplitudes involves a sum composed of two parts
of equal absolute value but opposite sign. Consequently, they are irrelevant
for the evolution procedure.\newline
The potential $V_{2}(\overline{r})$ proves invariant under interchange of
two coordinates. Its symmetry properties entail that only those amplitudes $%
p(\overline{r})$ contribute which in each direction involve equal positive
(bosonlike) parity under $P$ and $\widehat{P}$, i.e. 
\begin{equation}
\left. p_{\text{bbb}}^{\text{bbb}}(\overline{r})\right| _{V_{2}}\propto 
\hspace{-0.1in}\!\!\!\!\!\sum\limits_{P_{x},P_{y},P_{z},\widehat{P}_{x}%
\widehat{P}_{y},\widehat{P}_{z}}\!\!\!\hspace{-0.15in}\exp (-\tau
\,P_{x}P_{y}P_{z}\widehat{P}_{x}\widehat{P}_{y}\widehat{P}_{z}\,\overline{r})%
\stackrel{\tau \ll 1}{:\longrightarrow }\exp \left( -\tau V_{2}(\overline{r}%
)\right) \ \ \ .  \label{fac}
\end{equation}
Obviously, $p_{\text{bbb}}^{\text{bbb}}(\overline{r})$ is positive on the
reduced state space $\widehat{D}_{\,2}^{\,3}$. By recombination of both
potentials $V_{1}(\overline{r})$ and $V_{2}(\overline{r})$, the total
evolution scheme is established. Its proper implementation guarantees
accurate probabilistic sampling of the ortho-helium propagator on the
reduced state space $\widehat{D}_{\,2}^{\,3}$. The sampled density consists
of 4 subdensities $\Psi _{\text{fbb}}^{\text{fbb}}(\bar{r},\tau )$, $\Psi _{%
\text{ bfb}}^{\text{bfb}}(\bar{r},\tau )$, $\Psi _{\text{bbf}}^{\text{bbf}}(%
\bar{r},\tau )$ and $\Psi _{\text{fff}}^{\text{fff}}(\bar{r},\tau )$. Due to
the symmetry properties of the ortho-helium potential $V(\overline{r})=V_{1}(%
\overline{r})+V_{2}(\overline{r})$, certain classes of subprocess
transitions are forbidden; each of the subprocesses undergoes one out of
four possible transitions ruled by the probabilities $p_{\text{ffb}}^{\text{%
ffb}}(\overline{r})$, $p_{\text{fbf}}^{\text{fbf}}(\overline{r})$, $p_{\text{%
bff}}^{\text{bff}}(\overline{r})$ and $p_{\text{bbb}}^{\text{bbb}}(\overline{%
r})$. During a time step $\Delta \tau $, the subdensity $\Psi _{\text{fbb}}^{%
\text{fbb}}(\overline{r},\tau )$, for instance, branches into the
subdensities $\Psi _{\text{bfb}}^{\text{bfb}}(\overline{r},\tau +\Delta \tau
)$, $\Psi _{\text{bbf}}^{\text{bbf}}(\overline{r},\tau +\Delta \tau )$, $%
\Psi _{\text{fff}}^{\text{fff}}(\overline{r},\tau +\Delta \tau )$ and $\Psi
_{\text{fbb}}^{\text{fbb}}(\overline{r},\tau +\Delta \tau )$ according to
the probabilities $p_{\text{ffb}}^{\text{ffb}}(\overline{r})$, $p_{\text{fbf}%
}^{\text{fbf}}(\overline{r})$, $p_{\text{bff}}^{\text{bff}}(\overline{r})$
and $p_{\text{bbb}}^{\text{bbb}}(\overline{r})$. Notice that the analysis of
the subprocess transition amplitudes is consistent with the restriction to
four subdensities in the sense that no transitions occur to subdensities
different from the four specified ones. Fig. 6 illustrates the energy
estimates for the lowest ortho-helium state achieved with a pre-thermalized
sample. The numerical average -2.1744$\pm $0.0012 H is in good agreement
with the numerical energy estimate of -2.175229 H found in \cite{UMR88}.
Systematic errors due to time discretization and imperfect sampling could be
diminished by the use of total sample sizes of 100,000 walkers and time
steps of 0.001 H$^{-1}$.

\section{Discussion and conclusions}

It should be mentioned that the relation between fermi statistics and the
elimination of paths at permutation hyperplanes has been put forward by
Korzeniowski et al. \cite{KORZE}. This approach was subsequently criticized 
\cite{REPLY}, because it left open a few fundamental questions. Studying the
underlying mathematical Ansatz has lead us to an extension of the recently
reported many-body diffusion approach \cite{BDL94,BDL95,BDL96,BDLer97} in
which it was the basic idea to rigorously separate the multi-dimensional
free diffusion process of indistinguishable particles into one-dimensional
free diffusion processes. The novelty of the outlined symmetry analysis for
the simulation of indistinguishable particles is the construction of
numerically implementable diffusion process with a switching of boundary
conditions driven by a compound Markov chain . The continuous-time Markov
chain, not considered before in that framework, has been set up to enable
the realization of interdependent Brownian motions. The inclusion of
transitions between a set of simultaneously evolving diffusion processes has
been shown necessary to approach identical particles in potentials which are
not invariant under permutations of Cartesian coordinates. To transfer the
recently reported MBDA sampling technique \cite{LBDL97} to those systems, a
symmetry analysis of the potential is necessary. In applications to
different model systems, the consequences for the total diffusion process
have been demonstrated. The numerical findings confirmed the feasibility of
the utilized symmetry concept. The model of one particle exposed to a
double-well or a Mexican-Hat potential transparently illustrates the essence
of our approach. Artificially creating a sign problem by the decomposition
of the propagator in two resp. four subpropagators, some of which are not
positive definite on the configuration space, the efficiency of the
algorithm has been found comparable to the corresponding boundary-free
formulation. This application points out the generality of the symmetry
analysis on which the algorithm is founded; its use is not restricted to the
description of identical particles but may also be of relevance for the
implementation of boundary conditions on distinguishable particle systems.
In the MBDA, the boundaries guarantee the generation of symmetric and
antisymmetric Brownian motions. This is fundamentally different from nodal
approaches; obviously, the ground-state wave function of the particle does
not vanish at the boundaries (cfr. figs. 2 and 4)! The flexibility of the
many-body diffusion Ansatz has been indicated by treating two identical
fermion oscillators in three dimensions with an algorithm based on the same
principles but different subprocess-transition probabilities. Evidence has
been given that the MBDA avoids the fermion sign problem in applications to
the ortho-helium model. Necessary ingredients for this study have been the
derivation of the appropriate reduced state space $\widehat{D}_2^3$ with the
properly adapted stochastic processes. The comparison of the numerical
ground-state estimates indicated that the MBDA reliably simulates the
exchange contributions. Over the considered evolution intervals no
significant decay in the signal-to-noise ratio has been observed. The
reduced state space boundary $\partial \widehat{D}_2^3$ does definitely not
represent a nodal surface. Also in numerical practice, the characteristic
fermion state is completely determined by the Euclidean-time Schr\"{o}dinger
equation supplied with the symmetry properties of the free-diffusion process
and the potential under consideration. From that point of view, the MBDA
represents a potential candidate for ab-initio fermion Monte Carlo
simulations. In comparison with literature the MBDA---in its present
form---is restricted by its versatility. The MBDA tackles fundamental
methodological aspects to avoid the sign problem for fermions. To
transparently demonstrate that the supply of nodal surfaces is unnecessary
for the current approach, the present formulation dispenses with such
analytical assumptions. We are well aware that importance sampling may
drastically enhance computational efficiency, but the objective has been to
keep the algorithm as transparent as possible. The nodal structure of the
ortho-helium ground state, for instance, can be derived by symmetry
arguments \cite{KLE76}. Implementing an importance function with this nodal
structure enables one to circumvent the sign problem. The MBDA algorithm,
however, does not require to build in the nodal structure. We stress that
the MBDA is fundamentally different from approaches based on interacting or
correlated walker ensembles \cite{ARN82,AND91,ZHA91,LIU94,KAL96}. Dealing
with many-fermion diffusion on the appropriate state space $D$, the
ground-state wave function is simulated by a sum of antisymmetric elementary
densities. This is in contrast to the common definition of positive and
negative walker ensembles, the densities of which behave naturally
symmetric. The MBDA involves only positive walkers. Antisymmetry is realized
by establishing the correct free diffusion process without requiring
projection by an antisymmetric trial function. The use of reflection and
absorption restricts the evolution to the state space; this procedure proves
useful as long as the supplementary effort for reflection and absorption is
balanced by enhanced accuracy due to restriction to the state space. The
present investigations mainly attempted to solve and clarify the fundamental
difficulties related to the sign problem for fermions. The underlying
symmetry concepts were analyzed and used to guarantee the stable simulation
of fermion systems. Some transparent examples were chosen mainly for
illustrative purpose and no inference has been made to provide ultimate
accuracy.

\acknowledgments
\label{0references}Work supported by the ``PHANTOMS onderzoeksnetwerk'' and
Interuniversitaire Attractiepolen - Belgische Staat, Diensten van de Eerste
Minister - Wetenschappelijke, technische en culturele aangelegenheden. We
acknowledge support by the BOF NOI 1997 projects of the Universiteit
Antwerpen, the FWO projects G.0071.98 and WO.073.94N (Wetenschappelijke
Onderzoeksgemeenschap, Scientific Research Community of the FWO on
''Low-Dimensional Systems''). Part of this work has been performed in the
framework of the projects G.0287.95, 1.5.545.98 and the supercomputer
project 1.5.729.94N of the FWO. F.L. is very grateful to B. Gerlach for
partial financial support during the development of this work.

\begin{center}
{\sc Figure Captions}
\end{center}

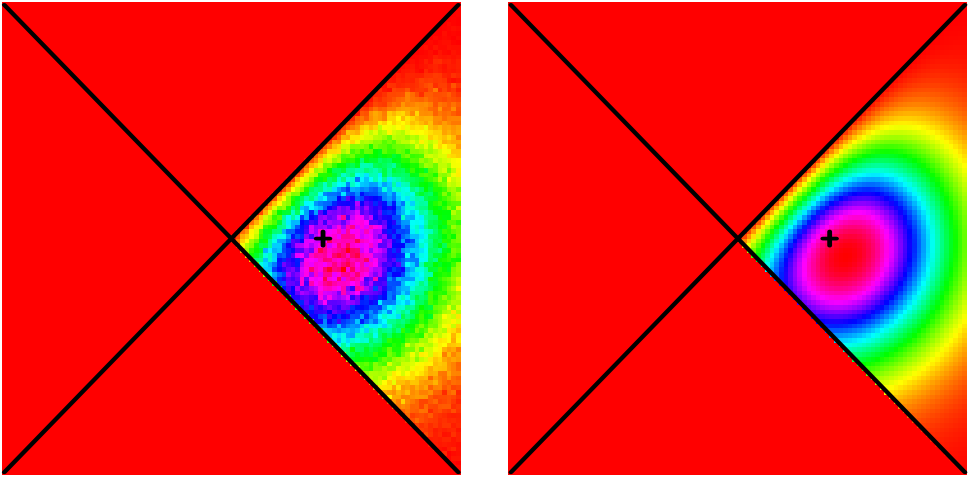
{\bf Fig. 1}: The rigorous (right hand side) and numerically sampled (left
hand side) Green function of two free identical fermions in one dimension
restricted to the reduced state space $x_1\geq |x_2|$ after an evolution of
three atomic time units. The initial condition is denoted by the crosses,
the state-space boundaries are indicated by the solid lines.

{\bf Fig. 2}: Plot of the symmetric and asymmetric double well potential $%
V_s(x)$ and $V_a(x)$ (\ref{double-well}) with $a=2$. Solid line: $b=0$,
dashed line: $b=0.25$. The long-dashed resp. dot-dashed curve depicts the
corresponding simulated ground-state wave functions $\Psi _s(x)$ and $\Psi
_a(x)$.

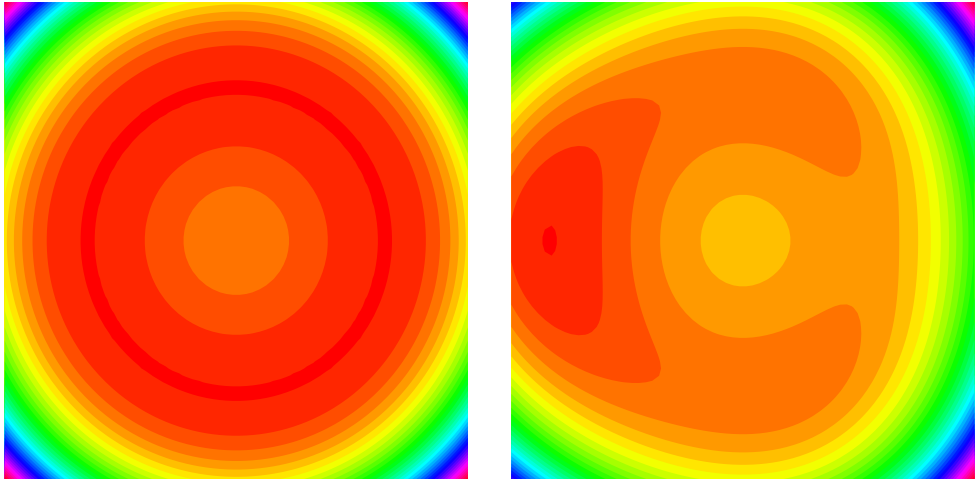
{\bf Fig. 3}: Plot of the Mexican Hat potential (\ref{MexiHat}) with $a=2$
and $b=0$ (l.h.s.) and with $a=2$ and $b=0.25$ (r.h.s.).

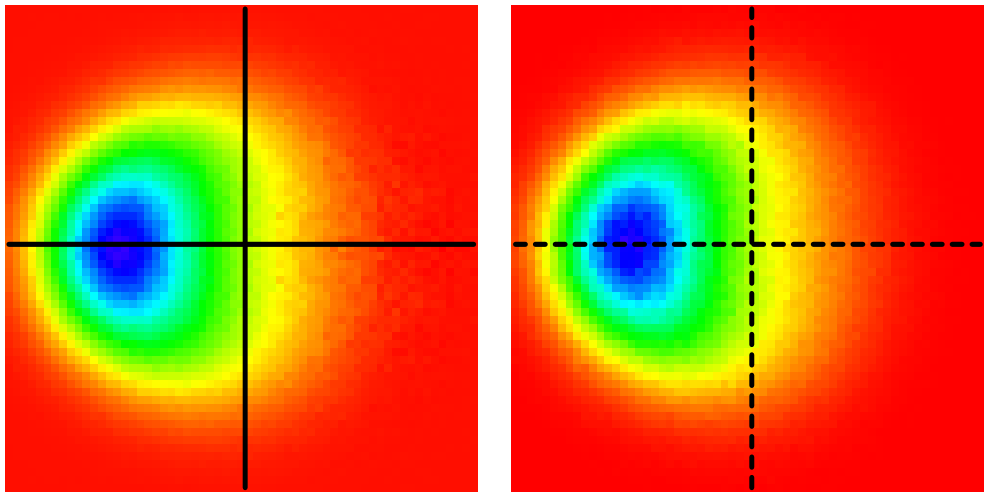
{\bf Fig. 4}: Comparison of the two numerically predicted ground-state
wavefunctions of the asymmetric Mexican-Hat model (\ref{MexiHat}), with $a=2$
and $b=0.25$ sampled with the MBDA (l.h.s) and the boundary-free algorithm.

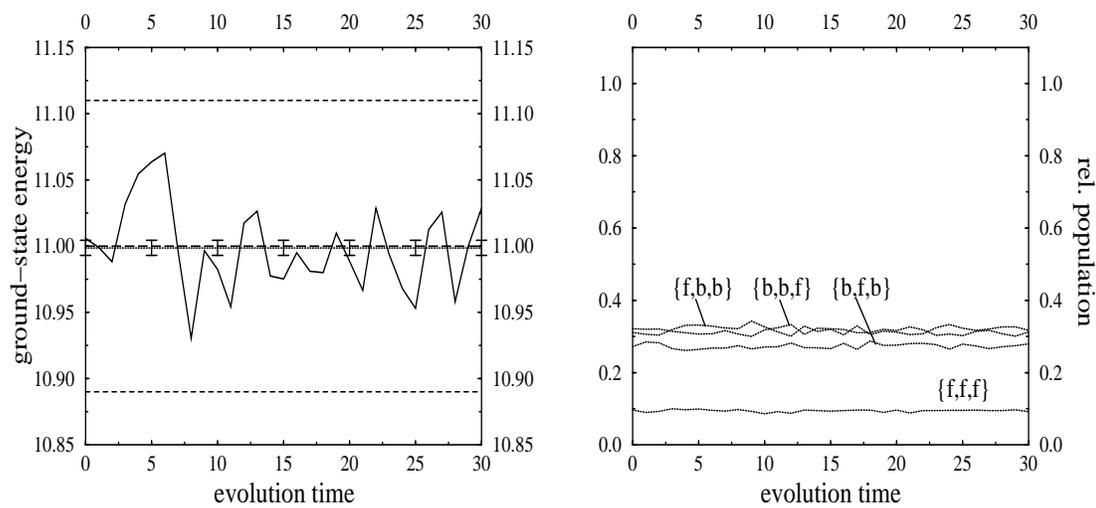
{\bf Fig. 5:} The ground-state energy estimates (left hand side) and the
relative walker population of the subdensities (right hand side) as a
function of Euclidean evolution time for the three-dimensional anisotropic
system of two identical fermion oscillators.

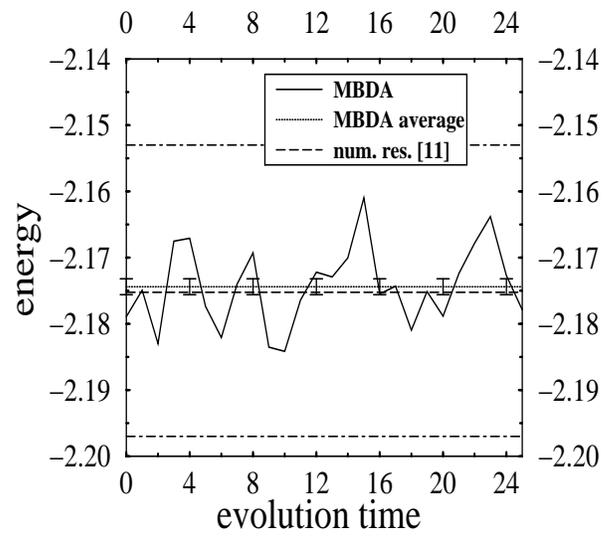
{\bf Fig. 6:} Comparison of the numerical energy estimates (solid curve) for
the lowest ortho-helium state with the numerical estimate -2.175229 H found
in \cite{UMR88} (dashed curve). The dot-dashed lines denote a deviation of
one percent from the dashed curve.

{\bf Tab. 1: } Estimated ground-state energies $E_0$ with standard
deviations $\sigma $ and relative subprocess population for the double-well
potential potential (\ref{double-well}).

{\bf Tab. 2:} Same as table 1 but for the Mexican-Hat potential (\ref
{MexiHat}). \bigskip

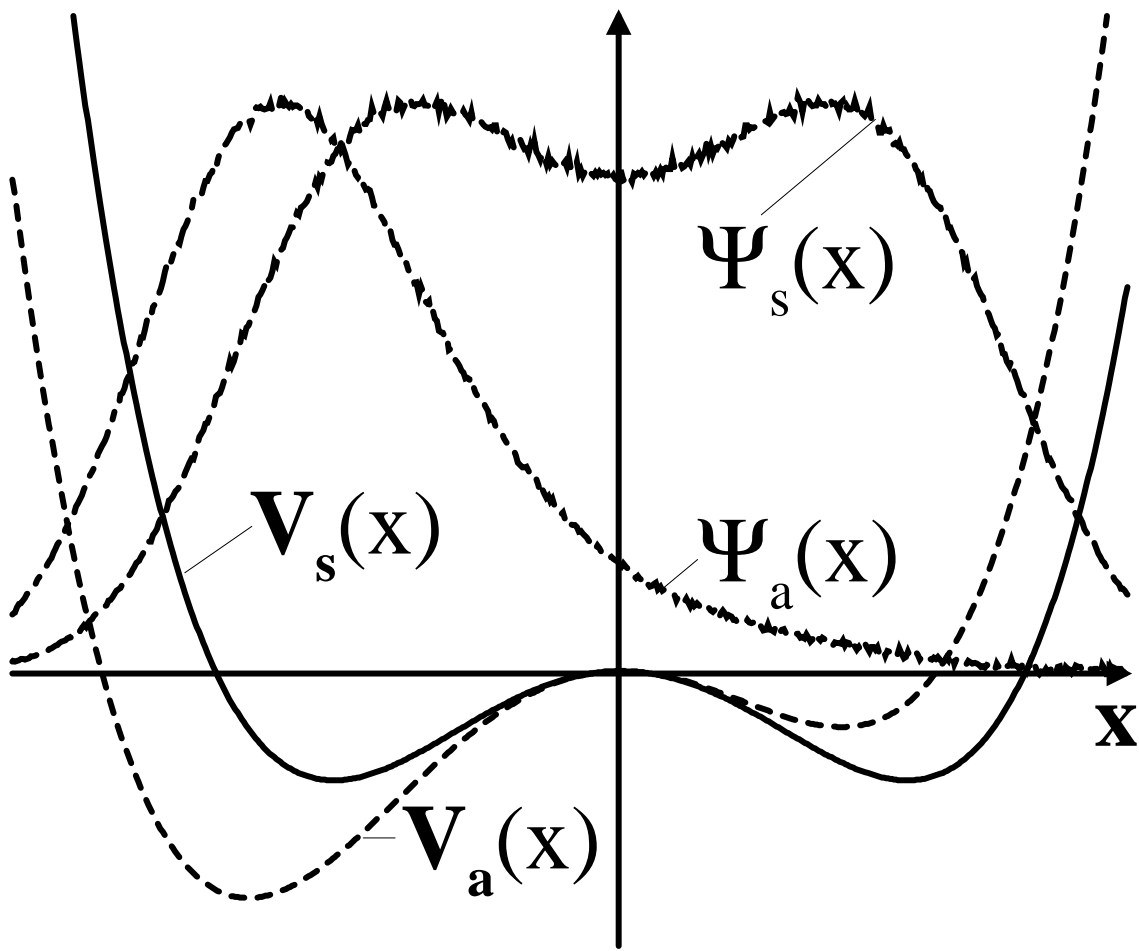
\end{document}